\documentclass[useAMS,usenatbib,referee]{mn2e}
\usepackage{epsf}
\usepackage{natbib}
\usepackage{graphicx}
\usepackage{epsfig,rotate}

\bibpunct{(}{)}{;}{a}{}{,}

\newcommand{\be} {\begin{equation}}
\newcommand{\ee} {\end{equation}}

\def\witchbox#1#2#3{\hbox{$\mathchar"#1#2#3$}}
\def\leqsim{\mathrel{\rlap{\lower3pt\witchbox218}\raise2pt\witchbox13C}}
\def\geqsim{\mathrel{\rlap{\lower3pt\witchbox218}\raise2pt\witchbox13E}}

\title
{On  the dynamics of a twisted disc immersed in a radiation field }
\author[P. B. Ivanov and
J. C. B. Papaloizou]{
P. B. Ivanov$^{1,2}$\thanks{E-mail:pbi20@cam.ac.uk
(PBI) J.C.B.Papaloizou@damtp.cam.ac.uk (JCBP)} 
and J. C. B. Papaloizou$^{1}$
\footnotemark[1]\\
$^{1}$Department of Applied Mathematics and Theoretical Physics,
University of Cambridge,\\
Centre for Mathematical Sciences, 
Wilberforce Road, Cambridge, CB3 0WA, UK \\
$^{2}$Astro Space Centre, P. N. Lebedev Physical Institute,
4/32 Profsoyuznaya Street,
Moscow, 117810, Russia}

\begin{document}

\date{Accepted Received ; in original form }


\maketitle


\begin{abstract}

We study the dynamics of a twisted tilted disc under the influence
of an external radiation field. Assuming the effect of absorption
and reemission/scattering  is that a pressure is applied to the disc surface where
the local optical depth is of order unity, we determine the response
of the vertical structure and the influence it has on the possibility
of instability to warping.  

We derive a pair of equations describing the evolution of a  small tilt
as a function of radius in the small amplitude regime that applies to 
both the diffusive and bending wave regimes. We also study the non linear
vertical response of the disc numerically using an analogous  one dimensional
slab model. 
For global warps, we find that in order for the disc vertical structure
to respond as a quasi uniform shift or tilt, as has been assumed
in previous work,  the product of the ratio
of the  external radiation momentum flux to the local
disc mid plane pressure, where it is absorbed,  with the disc aspect
ratio should be significantly less than unity. Namely, this quantity
should be of the order of or smaller than the ratio of the disc gas density
corresponding to the layer intercepting radiation to the mid plane density,
$\lambda \ll 1$. 

When this condition is not satisfied the disc surface tends to
adjust so that the local normal becomes perpendicular
to the radiation propagation direction. In this case dynamical quantities
determined by the disc twist and warp tend to oscillate with a large characteristic
period $T_{*}\sim \lambda^{-1}T_{K}$, where $T_{K}$ is some 'typical' orbital 
period of a gas element in the disc. The possibility
of warping instability then becomes significantly reduced.

In addition, when the vertical response is non uniform,
the  possible  production of shocks may lead to an important  
dissipation mechanism.

\end{abstract}

\vspace{-1cm}

\section{Introduction }
\noindent

A significant number of X-ray binaries  exhibit long-term  periodicities on time-scales of ~10-100 d. 
Examples are Her X-1, SS 433
and LMC X-4  see e.g. Clarkson et al. 2003.
 Precession and warping  of a tilted accretion disc
 has been proposed as an explanation (Katz 1973; Petterson 1975)
 which has subsequently  been  found
  to have observational support (Clarkson et al. 2003). 
The effect of radiation pressure  on a twisted tilted accretion
disc was first considered by Petterson 1977a,b who noted that 
when radiation
from the central source is absorbed at the disc surface and  re-emitted,
 a potentially important torque will result.
Iping \&
Petterson 1990 later 
suggested that such torques
determined the shape of the disc and its precession rate.  
Pringle 1996 subsequently showed that an initially axisymmetric thin disc could
be unstable to  warping as a result of interaction with an external radiation field
(see e.g.  Maloney et a. 1998; Ogilvie \& Dubus 2001 and Foulkes et al. 2006 for
later developments).
 
The analysis considered the disc to behave as a collection of rings,
which could interact by  transferring  angular momentum through the action of viscosity, but
otherwise behaved as if they were rigid. In particular, the vertical displacement
or tilt was assumed to be essentially uniform and independent of the vertical coordinate.
Thus the pressure applied at the surface at optical depth unity is assumed to be effectively
communicated through the disc vertical structure so as to give a near uniform response.
In this paper we extend the theoretical treatment of the interaction of a disc 
with an external radiation field to take account of possible significant departures of the tilt or displacement
from uniformity in the vertical direction. One of our objectives is to determine
the conditions under which the assumption of a uniform response is valid,
and then to estimate some of the consequences when they are not satisfied,
including potential additional dissipation resulting from non linear effects
such as the production of shock waves. The latter is done using a one dimensional
slab analogue model with the required high resolution in the vertical direction.
 
Although  we focus on the effects of a surface pressure
induced by an external radiation field, very similar considerations are likely
to apply when warps are induced by a surface pressure resulting from interaction with
an external  wind
(e.g. Quillen, 2001)
or surface forces produced by the interaction of the disc with an 
external magnetic field originating in the central star
( e.g. Pfeiffer \& Lai, 2004). 

We begin by considering the relevant issues using simple physical arguments. Following  Pringle 1996
we consider a thin disc immersed in an external radiation field
with mid plane initially coinciding with a
Cartesian $(x,y)$ plane. It is supposed that the upper surface
is parallel to the external rays so that there is initially
no interaction with the radiation.  The upper surface is then given 
a vertical elevation 
$h(r,\phi),$ where we now use polar coordinates $(r,\phi).$
 As a result, a disc element with
surface area $d{\cal A}$ absorbs  momentum from the radiation field
 at a rate
\be  {\dot {\cal F}} = F_0  d{\cal A} \left(r{\partial (h/r)\over \partial r}\right).\ee
Here $ F_0$ is the momentum flux at radius $r$ associated with the
external radiation field. The factor in brackets gives the angle through which
the local normal is rotated as a result of the perturbation (see Appendix \ref{A} for more details).
Assuming the absorbed momentum is reradiated isotropically above the disc
 by the surface layers at an  optical depth of unity,
there will be an applied pressure there of magnitude
\be  {\cal P} = {2F_0\over 3}   \left(r{\partial (h/r)\over \partial r}\right).\ee
This external pressure, when applied to a complete
elementary ring, produces a net torque per unit  length 
of magnitude
\be { d {\cal T} \over dr} = {2\pi F_0 \over 3} 
\left(r^2{\partial (h/r)\over \partial r}\right){\bf l},\ee
where the complex  vector ${\bf l}$ has  components equal to $(i,1,0)$ in the
Cartesian coordinate system. 
Here, in performing   the azimuthal integration,  we take into account that
the azimuthal dependence of $h$ is through a factor of the form 
$ \exp(-i\phi)$  
and work with the radial amplitude from now on.

To find the consequent evolution of the disc, one requires the component
of the  angular 
momentum per unit length perpendicular to its unperturbed direction.
This is given by

\be {d{\cal J}/dr} = 2\pi\Sigma r^2 \Omega {\langle h \rangle}i{\bf l},\ee
with $\Omega$ and $\Sigma $  being the near Keplerian local disc angular velocity and the
disc surface density, respectively. $\langle h \rangle =\int  d\zeta \rho h /\Sigma $ where
$\zeta $ is a vertical coordinate and $\rho $ is the gas density.

For an elementary ring the condition that the
rate of change of angular momentum equal the applied torque gives
a tilt evolution equation of the form
\be {\partial {\langle h \rangle}\over \partial t}=
-i{ F_0 \over 3\Omega\Sigma} 
\left({\partial (h/r)\over \partial r}\right). \ee
Assuming the vertical response is uniform,
so that we can set $h= {\langle h \rangle} \equiv r{\bf W}$,
we obtain a description of warp evolution equivalent to Pringle 1996
when effects due to viscosity and bending wave propagation are neglected 
(see sections \ref{simpa} and \ref{Dyneq} below). This description indicates
the possibility of instability to radiation pressure warping.

However, here we stress the fact that the averaged  elevation 
$\langle h \rangle ,$ enclosed in angled brackets,
applies to the bulk of the inertia of the disc and can
therefore be shown to be close to the elevation of the mid plane.
On the other hand $h$ as used in the torque formula applies to the
disc surface elevation. An important aspect of this paper is to distinguish
these two elevations and investigate under what conditions they
can be taken to be equal. This would be possible if the vertical structure
responds as a rigid body to the external pressure forcing and we find the
conditions for this to occur. The general requirement is found
to be that the density at the surface of the disc where the 
pressure is applied should not be too small.

It is possible to estimate in a simple way when the vertical displacement
response
of the disc to the external pressure forcing becomes non uniform.
Let us suppose that the external pressure $ {\cal P}$ is applied
at the surface layer where the density $\rho = \rho_*.$
The induced vertical displacement $h,$  now should be considered to  be
also  a function
of  the vertical coordinate $\zeta.$

Assuming a linear response, which should be appropriate for sufficiently
small elevations and, for simplicity an isothermal equation of state
with sound speed $c_s,$ it can be easily shown that in the upper layers 
of the disc where vertical motions dominate the Lagrangian pressure perturbation
has the form determined by presence of $h$, $\Delta P = -\rho_{*}c_{s}^{2} {\partial h\over \partial \zeta}$,
see equation (\ref{e24}) below. Equating this to the external pressure 
we have 
\be {\partial h\over \partial \zeta}= -{{\cal P}\over \rho_* c_s^2}.\ee   
If the vertical extent of the disc is $\zeta_0,$ and $h$ varies
on a radial scale comparable to $r,$  the characteristic change
in $h$ induced over the vertical thickness is easily estimated to be

\be \Delta h \sim  \zeta_0{{\cal P}\over \rho_* c_s^2} \sim {2F_0r \over 3\rho_* \Omega^2\zeta_0}
\left({\partial (h/r)\over \partial r}\right),\label{i1}\ee
where we use the approximate relation $c_s=\Omega\zeta_0.$ 

From the condition $\Delta h \sim h $, using (\ref{i1}) and estimating ${\partial h\over \partial r}\sim h/r$, 
it is clear that whether $h$ is uniform or not
 is governed by the  the parameter
$ F_0 r /(\rho_*  \Omega^2\zeta_0^3)(\zeta_0/r)^2 = \lambda^{-1} \epsilon \delta^2,$
with $\lambda =\rho_*/\rho_c$ and $\delta = \zeta_0/r$  so defining
 $\epsilon= F_0 r /(\rho_c  \Omega^2\zeta_0^3).$

For a uniform response,
one requires that $\epsilon \delta ^2 \ll \lambda.$
This is equivalent to the requirement that the product of the ratio
of the  external radiation momentum flux to the local 
disc pressure with the disc aspect 
ratio should be significantly less than unity.
We recall here that because of the geometrical configuration,  the  momentum flux 
locally reradiated by  the disc is in general much less than the external
momentum flux at that location.

In  this paper we extend the treatment of disc warping
induced by the action of an external pressure to include the effects of
the response of the vertical structure particularly when this is non
uniform as is the case when the above condition is not satisfied.
In that case we find that the disc dynamics enters a different regime.
This is such that the exposed surface tends to align so as to reduce the
momentum absorption and hence the applied pressure. In the extreme limit of this
regime the upper surface acts as if it is 
in contact with a rigid wall with the tendency to radiation warping instability tending to vanish.
In this limit quantities characterising the disc twist and warp tend to oscillate at 
typical frequency $\sim \lambda \Omega $.

To illustrate these effects we derive a description of the one dimensional evolution
of the disc inclination in radius and time that incorporates the effects of
the vertical structure response, radiation torques and  which applies both 
to the high viscosity regime,  
when the evolution is diffusive, and to the low viscosity regime when the evolution is wavelike.
In all cases, the efficacy of surface  radiation pressure driven instabilities
is found to be reduced once the model parameters are such that the vertical
response is significantly non uniform.

We go on to estimate conditions for the response to be nonlinear
and investigate the development of shock waves in the response
 using a one dimensional slab analogue model which has the same
 characteristic behaviour of  linear perturbations  
as the full disc model. We find that such shocks potentially  provide an
important dissipation mechanism.

The plan of this paper is as follows.
In section \ref{sec1} we describe some aspects of the thin disc model used.
In section \ref{Coords} we go on to introduce the twisted coordinate system
used  together with the notation convention, giving the basic equations in
section \ref {Beq}. By integrating over the vertical direction 
we use these to  derive  a single equation governing   the dynamics of
a twisted disc  in section \ref{Derveq}. When the disc behaves like
a set of rigid rings for which the vertical response is uniform,
this equation can be used to give a complete specification of the warp evolution.
We note that this provides an  extension of  previous formulations
 to be able to consider the case when  warps propagate
as waves rather than diffuse radially (e.g. Nelson \& Papaloizou 1999).

In this Paper we use the  twisted coordinate system  formalism 
first introduced by
Petterson 1977a, 1978 where the dynamical equations take the most simple form. When this formalism
is adopted and an accurate description of all components of the equations of motion is needed
as in the problem on hand we show that, in general, there is an ambiguity in choice of 
twisted coordinates with a set of these describing the same physical situation. As discussed in
section \ref{Derveq} a choice of a most appropriate twisted coordinate system can be
motivated by the condition that perturbations of all dynamical quantities determined by the
disc twist and warp are small. The transformation law between different twisted coordinates 
 corresponding to the same physical situation  is  derived in appendix C for an inviscid disc.

In order to obtain a complete description of the warp evolution
when the vertical response is non uniform,
we begin by  obtaining a complete solution of the vertical 
problem for a polytropic model in section \ref{poly}. This is used to obtain a pair of equations governing the 
radial warp evolution. We also indicate how the results can be extended to
apply to  more general models. We confirm the condition for the disc response
to be like that of a series of rigid rings as $\lambda \gg \epsilon (\zeta_0/r)^2.$

We go on to perform a linear stability analysis of the radial evolution equations
adopting a WKB approach in section \ref {qualan} obtaining 
a maximum potential growth/decay rate in section \ref{7.3}.
In section \ref{8} we discuss and confirm the analogy between the response of the disc
and the  linear and non linear dynamics of a vertically stratified 
one dimensional slab. We consider the 
development of shock waves and the formation of a rarefied hot atmosphere
in section \ref{8.3} giving a crude estimate of the warp dissipation rate
in section \ref{8.4}. Finally in section \ref {Conc} we discuss and summarize
our results.

\section{A thin disc model }\label{sec1}
\noindent To begin  we briefly summarize some of the properties of
an unperturbed axisymmetric thin accretion disc model which are
used in the discussion below.

However, the physical basis of our  results does not
depend on the detailed properties of this  model which has been
used for the purpose of making specific calculations. In order to
make our treatment as simple as possible we use a highly
simplified version of the Shakura $\&$ Sunyaev 1973 $\alpha
$-disc model. We assume a polytropic equation of state for the
disc gas \be P=K(r)\rho^{\gamma }     \label{e1}\ee where $P$ and
$\rho $  are the gas pressure and density, $r$ is the radial
coordinate and $\gamma $ is the assumed constant specific heat
ratio. It is assumed that the  "polytropic constant" $K$ is, in
general, a function of the  radial coordinate, $r,$ but  does not
depend on the local vertical coordinate $\zeta $, see  equations
(\ref{e7}), (\ref{e8})  below for a definition of the coordinate
system used in our study. The equilibrium variation of density and
pressure with height follows from the equation of state (\ref{e1})
together with  the equation of hydrostatic equilibrium \be
{\partial P \over \partial \zeta}=c_{s}^{2}{\partial \rho \over
\partial  \zeta} =-\rho \Omega^{2}\zeta . \label {e2}\ee Here
$c_{s}$ is the adiabatic sound speed, $\Omega =\sqrt {(GM)/
r^{3}}$ is the Keplerian angular velocity and $M $ is the mass of
a central object.

Integration of equation (\ref{e2}) gives
$$ \rho =\rho_{c}(r)\left(1-\left({\zeta \over \zeta_{0}(r)}\right)^{2}\right)^{{1\over
1-\gamma}}, \quad P =P_{c}(r)\left(1-\left({\zeta \over \zeta_{0}(r)}\right)^{2}\right)^{{\gamma \over
1-\gamma}},$$ 
\be \qquad \qquad \qquad \qquad \qquad \qquad c_{s}^{2}=\gamma {P_{c}\over \rho}
\left(1-\left({\zeta\over \zeta_{0}(r)}\right)^{2}\right),   \label{e3}\ee
where $\rho_{c}$ and $P_{c}$ are the mid plane values of the density
and pressure, respectively and $\zeta_{0}$ is the disc semi-thickness.
These quantities are related to each other  through
\be \Omega^{2}\zeta^{2}_{0}={2\gamma \over \gamma -1}{P_{c}\over
\rho_{c}}. \label{e4}\ee
As  shown below, many of  our equations take an especially simple form
for the case $\gamma =5/3,$ accordingly we adopt  this value below.

It is assumed below that the standard Navier-Stokes equations are
valid as the dynamical equations of our problem with the viscosity
law set up according to the standard prescription
(Shakura $\&$ Sunyaev 1973 )
\be \nu =\alpha \Omega^{-1}P, \label{e5}\ee
where $\nu $ is the kinematic viscosity. For
simplicity we assume below that $\alpha $ is a small constant parameter and
consider only leading terms in corresponding expansion series.

The  radiation intercepted by the disc is considered to result in
an external pressure applied to a surface near to the free surface
of the disc. This surface is formally defined through the condition
$\rho=\rho_{*}(\tau=1)$ where $\tau $ is the  optical  depth.
It is assumed hereafter that the ratio of  the density at this surface
to  the midplane density $\rho_{*}/\rho_{c} \ll 1$.
Thus, we neglect the effect of heating of the upper disc layers by incident
radiation. Also, for simplicity it is assumed that the radiation
does not affect the structure of  the unperturbed disc. This can be achieved by
setting the
disc semi-thickness to be proportional to the radial distance
\be \zeta_{0} =\delta r , \label{e6}\ee
where $\delta \ll 1$ is the disc opening angle which is constant.

\section{Coordinate system and notation convention}\label{Coords}

The dynamical equations governing a twisted disc take their
simplest form in the so-called twisted coordinate system first
introduced by Petterson 1977a. Since this system has been
discussed extensively elsewhere (for details see e.g. Petterson 1977a; 1978),
 in this paper we give only some basic definitions and
relations important for the future discussion.

In order to define the twisted coordinates let us consider
a fixed Cartesian coordinate system $(x_1,x_2,x_3)$ with
origin at the central star and
$(x_1,x_2)$ plane  coinciding with the mid-plane of the unperturbed disc.
We then perform two  elementary rotations of
the coordinate axes
parametrised by the  two Euler angles $\gamma $ and
 $\beta.$ The first rotation is of the
 $(x_1,x_2)$ axes through  the angle $\gamma$ keeping the
 $x_3$ axis fixed. As a result of
this rotation the  $x_1$ and $x_2$ axes
change their directions. The second rotation is of the
$(x_2, x_3)$ axes in the plane
perpendicular to the new fixed direction of the $x_1$ axis and is through the angle
$\beta.$ The
cylindrical coordinates $(r, \psi, \zeta)$ defined with respect to the
rotated axes now determine the twisted coordinate system.

Formally, in the linear approximation that the inclination $\beta $
is small, the transformation to the twisted coordinate system
$(r, \psi, \zeta )$ from the initial Cartesian coordinates
$(x_1,x_2,x_3)$ is given by
\be \left( \matrix{
r\cos \psi \cr
r\sin \psi \cr
\zeta \cr }\right) = B(\beta, \gamma)
\left( \matrix{
 x_1 \cr
 x_2 \cr
 x_3\cr} \right), \label{e7} \ee
where $B$ is the rotation matrix
\be B(\beta, \gamma )= \left( \matrix{
 \cos \gamma & \sin \gamma & 0 \cr
 -\sin \gamma & \cos \gamma & \beta  \cr
 \beta \sin \gamma & -\beta \cos \gamma & 1 } \right) ,\label{e8} \ee and
$\beta(t, r)$ and $\gamma(t, r)$ are, in general, functions of $r$ and
the time $t.$

Instead of the original Euler angles it is useful to
introduce  the combinations $\psi_{1}=\beta \cos \gamma $,
$\psi_{2}=\beta\sin \gamma.$ Further, we
define  the angle $\phi =\psi +\gamma$
which is used to replace the angle $\psi.$ We are thus able to consider the case
$\beta (r,t) =0$ for which the transformation (\ref{e8}) is
degenerate. The rotation matrix giving the transformation from the twisted coordinates back to the
Cartesian coordinates is given by transpose of $B$,
$B^{T}$. The cylindrical coordinates $(r_{c},\phi_{c},z_{c})$
 associated with the original Cartesian system are defined through:
 $x_{1}=r_{c}\cos \phi_{c}$,
$x_{2}=r_{c}\sin \phi_{c}$ and  $x_{3}\equiv z_{c}.$
From equations (\ref{e7}) and (\ref{e8}) we obtain
\be \sin \phi_{c} =\sin \phi  -{\beta \zeta \over r} \cos \phi
\cos (\phi - \gamma), \label{e8a}\ee
\be r_{c} =r -\beta \zeta \sin (\phi - \gamma ), \label{e8b}\ee
and
\be z_{c}= \zeta +\beta r \sin (\phi - \gamma ). \label{e8c}\ee
From equations (\ref{e8a}) and (\ref{e8b}) it follows that when $\zeta
=0 $ we have $\phi_{c}=\phi $ and $r_{c} = r .$
In addition, the difference $\phi_{c}-\phi $ is also first order in the
small angle $\beta.$ Thus we can set $\phi_{c}=\phi$ in equations for quantities
that are first order in $\beta.$

All vectors and tensors appearing in our
calculations are projected onto a local orthonormal basis
 ${{\bar e}}_{r}$, ${{\bar e}}_{\psi}$,
and ${{\bar e}}_{\zeta}$ (Petterson 1977a, 1978) where  the bar is
used to denote a vector quantity  from now on. The explicit
expressions for the above  basis vectors  are not important for
our purposes and can be found elsewhere, see e.g. Petterson
1977a, 1978, Demianski $\& $ Ivanov 1997 hereafter DI. We
note, however, that the basis can be uniquely defined by the
requirement that ${{\bar e}}_{\zeta}$ is the coordinate vector
such that ${{\bar e}}_{\zeta}\cdot \nabla \equiv {\partial \over
\partial \zeta}$ for $\zeta \rightarrow 0.$ All other vectors can
be found from the orthonormality conditions together with the
requirement that the coordinate axes form a right handed system.

\subsection{State variables}
Since the coordinate lines of the twisted coordinate system are
not orthogonal and both the coordinate lines and directions of the
basis vectors depend on time and on the radial coordinate,
differentiation of  quantities is best performed with help of
appropriate connection coefficients. Their explicit form can be
found in Petterson 1977a, 1978 and  DI. Also, we note that the
projections of the  velocity vector ${\bar { v}}$ onto the basis
vectors are not in general proportional to the time derivatives of
the appropriate coordinates of a particular fluid element, see
Petterson 1978. Accordingly, we give relations between these
components and the time derivative of the appropriate gas element
coordinates where this is explicitly needed.

Writing them in the twisted coordinate system, in the
approximation that the inclination angle
is small enough that a linearization procedure
in which quadratic and higher powers of $\beta$ can be neglected,
the Navier-Stokes
equations and the continuity equation then lead to:

\noindent  1) a set of equations for unperturbed
quantities $Q_{0}$ formally
coinciding with the equations describing them
for flat disc accretion.

\noindent 2) A set of equations for quantities $Q_{1}$,
which are non axisymmetric perturbations to $Q_{0}$  induced by
deformation of the disc shape.
The dependence of the latter quantities on $\phi$ is harmonic:
$Q_{1}=Q_{1}^{(1)}\cos \phi +Q_{1}^{(2)}\sin \phi $
\footnote{Strictly speaking, in order to satisfy boundary
conditions at the disc surface
we can assume a harmonic dependence of some of
these quantities only over a limited domain of the angle $\phi$, see
Appendix \ref {A} for details.
However, as follows from the results provided
there,  we can formally proceed assuming such quantities
depend harmonically on $\phi $ over the appropriate domain  and then
extend to  the whole  $\phi $ domain using the rule
discussed in  Appendix \ref{A}.}

It is  convenient to represent these perturbations
as well as the angles $\beta $ and $\gamma $ using the complex
quantities, $ {\bf Q}$ and ${\bf W},$ which are defined through
\be {\bf Q}=Q_{1}^{(1)}+iQ_{1}^{(2)}, \quad {\bf W}=\beta e^{i\gamma}=\Psi_{1}
+i\Psi_{2}. \label{e9} \ee

Here we note that we adopt the convention that
quantities in bold  represent complex perturbations.
This convention applies also to the components of a vector with
vectors themselves denoted with an over bar. When real
as in the perturbed equations of motion
(\ref{e11a})-(\ref{e15}) given below, these
are not in bold. Note also that for convenience we use $\rho_1$ and $\xi^{\zeta}$ 
to denote the complex density  perturbation and vertical component of
the  Lagrangian displacement below, and these are not in bold.

\section{Basic equations}\label{Beq}

As we mentioned above we write each of the variables entering  the
continuity and the Navier-Stokes equations as the sum of an  unperturbed part,
formally coinciding with a solution of the equations for axisymmetric thin  disc
accretion, and a perturbation.
Accordingly, we have
\be  v^{i} =  v_{0}^{i} +  v^{i}_{1}, \quad \rho
=\rho_{0}+\rho_{1}, \label{e10}\ee
for the components of velocity and the gas density respectively.
But as the only non zero component of the unperturbed
velocity is in the azimuthal direction and corresponds to Keplerian
rotation which we deal with explicitly,
we shall drop the subscript $1$ from the velocity perturbation.

The unperturbed gas density $\rho_{0}$ is given by equation (\ref{e3}).
We assume that only circular Keplerian motion of the gas is present in the
unperturbed state and accordingly have $ v_{0}^{\phi} =r\Omega
=\sqrt{{GM\over r}}$. Thus we neglect the radial component of the
unperturbed velocity $ v_{0}^{r}$ present in the thin disc mainly
due to action of viscosity. This can be done in the limit of
a sufficiently small value of $\alpha < 1$.

Apart from two significant modifications, our equations for the perturbed
quantities follow from the corresponding equations
given in  DI provided that terms involving quadratic and higher order
corrections in the assumed small parameter $\alpha $ are discarded and that the gravitomagnetic force in the
$\zeta $~component as well as
relativistic  post-Newtonian corrections to the
$r$ and $\phi $-components of the Navier-Stokes equations are neglected.

The most important modification stems from the fact that in the
present study an accurate description of vertical motions and
displacements is needed and therefore, we take into account all
contributions to the vertical velocity, ${v}^{\zeta}.$ These
consist of the contribution determined by time-dependent evolution
of the basis vector ${\bar {e}}_{\zeta}$ and the contribution
determined by time dependent vertical displacements of gas
elements in the disc. Accordingly, we have \be  {v}^{\zeta} =r{\cal
U} +v^{\zeta}. \label{e11}\ee where ${\cal U}=\dot \beta \sin
\psi  - \beta \dot \gamma \cos \psi =\dot \Psi_{1} \sin \phi -
\dot \Psi_{2} \cos \phi $ and the dot stands for the time
derivative. The relation of $v^{\zeta}$ to the corresponding
component of the  displacement vector $ {\bar \xi},$
$\xi^{\zeta},$ is given by  Petterson 1978 \be v^{\zeta} =\dot
\xi^{\zeta} + \Omega {\partial \over \partial \phi} \xi^{\zeta}.
\label{e16}\ee
 Only the first
term in (\ref {e11}), which arises from the time dependence of the
changing coordinate system, is normally taken into account in
studies of twisted discs that use the formalism of the twisted
coordinate system, see Hatchett at al 1981 and DI. However, the
role of the second term, which describes the disc motion relative
to the moving coordinates, is  essential for our purposes.

We also take into account the time derivative of the perturbed
density $\rho_{1}$ in the continuity equation and the time derivative
of $v_{\zeta}$ in the $\zeta$-component of the Navier-Stokes equation.
These terms are proportional to a small parameter $1/(\Omega
t_{ev})$ where $t_{ev}$ is a characteristic evolution time of the
twisted disc. That is for any perturbation quantity ${\bf Q},$ we can write
$|{\dot {\bf Q}}/{\bf Q}| = O(1/(\Omega
t_{ev})).$ However, even though they are small, in a formal sense,
when compared with leading terms in
a series expansion, when retained, they allow us to regularise some
otherwise formally singular
expressions, that arise when dealing with the particular problem of
calculating the vertical displacements
in the disc, see e.g. equation (\ref{e34}) below.

The perturbed Navier-Stokes equations can be divided into 'horizontal'
part incorporating the $(r)$ and $(\phi )$-components and 'vertical' part
corresponding to the
$(\zeta)$ component. The horizontal part follows from the equations provided
in DI  when the relativistic corrections are
neglected and the limit of small viscosity is adopted.
The  $(r)$-component can be written  as
\be \rho_{0}(\dot {{ {v}}^{r}}+\Omega (
{\partial \over \partial \phi} { v}^{r}
-2{ v}^{\phi}))=-\rho_{0}\Omega^{2}r\zeta {\cal W}-
{{\partial }\over {\partial \zeta}}t^{r\zeta}.\label{e11a}\ee
where ${\cal W}=\beta^{\prime }\sin (\psi ) - \beta \gamma^{\prime } \cos \psi
=\Psi_{1}^{\prime }\sin \phi - \Psi_{2}^{ \prime } \cos \phi $
and
\be t^{r\zeta}=-\eta {{\partial }\over {\partial \zeta}} {v^{r}}
\label{e12}\ee
is the $(r,\zeta)$-component of the viscosity tensor. In the expression for ${\cal W},$
a prime denotes differentiation with respect to $r.$

Differentiating the $(\phi)$-component of perturbed Navier-Stokes
equations with respect to $(\phi)$ we obtain
\be \rho_{0}{{\partial }\over {\partial \phi}}{\dot { v^{\phi}}}+
{\rho_0\Omega\over 2}
 \left({{\partial }\over {\partial \phi}}{ v}^{r}-
2{ v}^{\phi}\right)=
-{{\partial^2 }\over {\partial \zeta \partial \phi}}
t^{\zeta \phi},\label{e12a}\ee
where
\be t^{\phi \zeta}=-\eta {{\partial }\over {\partial \zeta}}{ v}^{\phi}.
\label{e13}\ee

The set of equations (\ref{e11}) and (\ref{e12}) can be used in order
to express  the perturbed velocity components in terms of the quantity
$\cal W$. As seen from equations (\ref{e11}) and (\ref{e12}), to the leading
order in small parameters $\alpha $ ( or equivalently $\eta$) and $1/(\Omega
t_{ev})$, the perturbed velocity components enter in  both equations
in the same combination
\be {{\partial }\over {\partial \phi}}{ v}^{r}-
2{v}^{\phi}.\label{e13a}\ee
Therefore, these equations are degenerate to leading order and
accordingly the next order corrections have to be
retained  (Papaloizou $\&$ Pringle 1983, hereafter PP). This leads
to an inverse dependence of the 'horizontal' part of the perturbed
velocity on the small parameters such that  both $ { v^{r}}$ and
${v^{\phi}}$ are $ \propto \min (\alpha^{-1}, 1/(\Omega t_{ev}))$
(see PP for details).

The $(\zeta )$-component of the Navier-Stokes equation is written in the form
\be \rho_{0}\left(\dot v^{\zeta} +\Omega {\partial \over \partial \phi} v^{\zeta}
+2\Omega r{\partial \over {\partial \phi}} {\cal U}\right)=
-\Omega^{2}\zeta \rho_{1} -{\partial \over \partial \zeta}P_{1}, \label{e14}\ee
where $\rho_{1}$ and $P_{1}$ are the perturbations of density and
pressure, respectively. These are related through $P_{1}=c_{s}^{2}\rho_{1}$
with the sound speed given by equation (\ref{e3}).

The continuity equation is formally equivalent to perturbed continuity
equation written for the case of a flat disc in cylindrical
coordinate system
\be \dot \rho_{1}+ \Omega {\partial \over \partial \phi}\rho_{1}+
{{\rho_{0}}\over r}{{\partial }\over {\partial \phi}}{ v}^{\phi} +
{1\over r}{{\partial }\over {\partial r}}
(\rho_{0}{ v}^{r}r)
+{{\partial }\over {\partial \zeta}}(\rho_{0}v^{\zeta})=0. \label{e15}\ee
We recall that the relation between the velocity
component $v^{\zeta}$ and the corresponding component of the
Lagrangian displacement vector $ {\bar \xi}$
is given by equation (\ref{e16}).

Apart from the terms proportional to $\cal W$ and $\cal U$
equations (\ref{e11a}-\ref{e15}) are formally equivalent to the corresponding
equations that would be written down directly in a fixed cylindrical coordinate system
$(r, \phi, \zeta)$ (PP).

The term proportional to $\cal W$ in equation (\ref{e11a}) is
essentially a projection of the pressure gradient. Due to the
presence of perturbations with odd symmetry with respect to $\zeta $
in twisted discs,  surfaces of constant pressure do not in general
coincide with the orbital planes of gas elements. Therefore, there
is is a projection of the  pressure gradient onto such an  orbital plane
which is described by this term, see Fig. 4 of Ivanov $\&$ Illarionov
1997 for graphic representation of this effect.

The term proportional to $2{\partial \over {\partial \phi}} {\cal U}$
in equation (\ref {e14}) is made up from  two
contributions. The first is related to the non-holonomic character
of our basis vectors.  As we have mentioned above in a non-stationary
situation, this leads to a non-zero vertical velocity even for a gas element being at rest with
respect to the twisted coordinate system, see equation (\ref{e11})
above. This fact  was noted for the first time by
Hatchett, Begelman $\&$ Sarazin 1981. The second contribution is
determined by non-inertial effects. These are taken into account by an
appropriate connection coefficient in the formalism developed by
Petterson 1978. Both contributions have exactly the same form. This
accounts for the factor $2$ in the expression $2{\partial \over
{\partial \phi}} {\cal U}$ in equation (\ref {e14}).

The set of equations (\ref{e11a}-\ref{e15}) can be brought into a
simpler form with help of the complex notation introduced through equation
(\ref{e9}) for the  density perturbation and the perturbations to the velocity components.
 Also, we assume hereafter that all perturbed quantities depend
on $\phi$ and time only through an exponential factor. Thus
\be {\bf Q }\propto e^{-i\omega t-i\phi}, \label{e17}\ee
where the frequency $\omega $ is, in general, a complex constant.

Subtracting equation (\ref{e12a}) from equation (\ref{e11a}) and
introducing the complex notation it is easy to see that the result
 contains the components of the velocity perturbation only in
the combination ${\bf v}_{+}={ {\bf v}}^{r}+2i{ {\bf v}}^{\phi}$.
The resulting equation  can thus be represented as \be i\omega
\rho {\bf v}^{+}+{\partial \over \partial \zeta}\eta {\partial
\over \partial \zeta}{\bf v}^{+} = i\rho \Omega^{2}r\zeta
{\partial {\bf W}\over \partial r},  \label{e18}\ee where we use
equations (\ref{e12}) and (\ref{e13}) and set $\rho\equiv
\rho_{0}$ from now on. The quantities ${ {\bf v}}^{r}$ and ${ {\bf
v}}^{\phi}$ can be separately expressed in terms of ${\bf v}_{+}$
taking into account the fact that in the leading approximation in
the small parameters we can equate expression (\ref{e13a}) to
zero\footnote{Setting (\ref{e13a}) to zero physically reflects the
fact that trajectories of gas elements in the disc may be
approximated as Keplerian ellipses in the leading order. Next
order correction determine parameters of the these ellipses and
effect of slow precession of their main axes on time scale of
order of $t_{ev}$, see DI and Ivanov $\&$ Illarionov 1997 for more
details on physical picture of the horizontal motions in the disc.
Formally, parameters and precession rate of these ellipses can be
obtained from equation (\ref{e18}).} and, accordingly, obtain \be
{{\bf v}}^{r}-2i{ {\bf v}}^{\phi}=0, \quad { {\bf v}}^{r}={1\over
2}{\bf v}^{+}, \quad  { {\bf v}}^{\phi}= -{i\over 4}{\bf v}^{+}.
\label{e19}\ee Now we can express  ${ {\bf v}}^{r}$ and ${ {\bf
v}}^{\phi}$ in terms of ${\bf W}$ with help of equations
(\ref{e18}) and (\ref{e19}). To do this we remark that the $\zeta$
dependence is dealt with by noting that  $\eta \propto P$ and the
solution is such that both ${{\bf v}}^{r}$ and ${ {\bf v}}^{\phi}$
are $\propto \zeta.$ We then use the resulting expressions to
eliminate these velocity components in the continuity equation
(\ref{e15}) so obtaining \be (\omega +\Omega ){\bf
\rho}_{1}=-i\left(\rho {\bf A} \zeta + {{\partial }\over {\partial
\zeta}}(\rho {\bf v}^{\zeta})\right), \label{e20}\ee where \be
{\bf A}={GM\over r^{1/2}}\rho^{-1}{{\partial }\over {\partial
r}}\left({\rho r^{-3/2}\over {\hat \omega }}{\partial {\bf W}\over
\partial r}\right), \label{e20a}\ee and \be \hat \omega = \omega
+i\alpha \Omega. \label{e20b}\ee Finally, equations (\ref{e14})
and (\ref{e16}) can be brought in the form \be -i\rho (\omega
+\Omega ){\bf v}^{\zeta}=2i\rho \Omega \omega r{\bf W}
-\Omega^{2}\zeta{\bf \rho}_{1}- {{\partial }\over {\partial
\zeta}}c_{s}^{2}{\bf \rho}_{1}, \label{e21}\ee and \be {\bf
v}^{\zeta}=-i(\omega + \Omega){\bf \xi}^{\zeta}. \label{e21a}\ee

\section{Derivation of a single equation describing  the dynamics of
a twisted disc}\label{Derveq}

At first let us point out that the hydrodynamical equations as
written in a twisted coordinate system cannot in principle
provide a single dynamical equation for ${\bf W}$ without
some additional considerations.
Indeed, since the Euler angles $\beta $ and $\gamma $ are
considered as dynamical variables in this formalism,
the total number of dynamical variables is
always larger by two than the number of equations actually 
needed to determine the dynamics of the system.

For example, in the case of a baratropic equation of state
$P=P(\rho ),$ there are six dynamical variables used in the
description of the problem, these being the  three components of
velocity, and the  density together with the  two Euler angles.
However the number of dynamical equations  available is  four,
these being the three components of the Navier-Stokes equations,
and the continuity equation. Since the number of equations is
smaller than the number of variables, there are in principal,
different sets of variables ${\bar v}(t, r)$, $\rho_{1}(t,r)$ and
${\bf W}(t,r)$ describing the same dynamical system and connected
to each other by certain transformation laws. These transformation
laws can be easily found from the condition that the  velocity and
density fields measured in some inertial coordinate system remain
unchanged when the set of dynamical variables associated with the
twisted coordinate system is transformed. Explicit expressions for
these transformation laws are given in Appendix \ref{C}. It is
interesting to note that the dynamical variables in the inertial
coordinate system play the role of gauge independent quantities
and the laws of transformation between different sets of variables
in the twisted coordinate system may be regarded as gauge
transformations. This situation has many analogies in different
physical systems, say in systems governed by General Relativity,
such as e.g. , the theory of small cosmological perturbations, see
e.g. Landau $\& $ Lifshitz 1975.

In order to begin the process of obtaining a single equation
 describing the dynamics of a twisted disc
referred hereafter as a 'dynamical equation' we substitute the density
perturbation given by equation (\ref {e20}) into the 'vertical' equation
 (\ref {e21}) thus having
\be {{\partial }\over {\partial \zeta}} \left(c_{s}^{2}\rho_0
{{\partial }\over {\partial \zeta}}{\bf v}^{\zeta}\right) +2\rho_0 \Omega \omega
{\bf v}^{\zeta}=-2\rho_0 \Omega^{2}\omega r{\bf W}-\rho_0 \Omega^{2} {\bf A}\zeta^{2}-
{{\partial }\over {\partial \zeta}}(c_{s}^{2}\rho_0 {\bf A}\zeta), \label{e22}\ee
where we use the hydrostatic balance equation (\ref{e2}) and approximate $\Omega
  +\omega \approx \Omega $ in the first term on the right hand
side. Equation (\ref{e22}) plays an important role in our
analysis. A dynamical equation follows from  (\ref{e22}) provided
that  appropriate boundary conditions at the disc surface are specified.

\subsection{A simple approach to  the
derivation of a dynamical equation}\label{simpa}

Before discussing an approach to the problem which fully takes into account
the vertical structure of the disc, we would like to show how to obtain
a governing dynamical equation from equation (\ref{e22}) in a
straightforward way. Although this approach is incomplete it
allows us to obtain a dynamical equation which is approximately
correct provided that the ratio of the density where the external
pressure is applied to the central density, $\lambda = \rho_{*}/\rho_{c}$,
is not too small.

To do this we integrate equation (\ref{e22}) over $\zeta$ taking into
account  the fact that the last term on the right hand side vanishes when $\zeta
\rightarrow \pm \zeta_{0}$ and can be discarded.
We thus obtain
$$ \left(c_{s}^{2}\rho {{\partial }\over {\partial \zeta}}{\bf v}^{\zeta}
+c_{s}^{2}\rho {\bf A}\zeta
 \right){\huge |}^{\zeta \rightarrow \zeta_{0}}_{\zeta \rightarrow
  -\zeta_{0}}+2\Omega \omega \int^{+\zeta}_{-\zeta}\rho {\bf v}^{\zeta}d\zeta = $$
\be \qquad \qquad \qquad \qquad \qquad -2\Sigma \Omega^{2}\omega r {\bf W}-{GM\Omega^{2}\over 2r^{1/2}}
{{d }\over {d r}}\left({\Sigma H^{2} r^{-3/2}\over
 \hat{ \omega }}{{d }\over {d r}}{\bf W}\right), \label{e23}  \ee
where $\Sigma =\int^{+\zeta_{0}}_{-\zeta_{0}}d\xi \rho_0$ is the surface density and
a typical disc height $H$ is defined by condition
$H^{2} =\int^{+\zeta_{0}}_{-\zeta_{0}}d\zeta \zeta^{2}\rho_0/\Sigma$.

It can be easily shown that the surface terms on the left hand side
are, in the low frequency limit, proportional to the
Lagrangian  pressure perturbation, $\Delta P=(dP_0 /
  d\zeta)\xi^{\zeta}+P_{1}$. Introducing the complex notation and
taking into account that in the low frequency limit, we   can
write from (\ref{e21a}) that ${\bf \xi}^{\zeta}\approx
i\Omega^{-1}{\bf v}^{\zeta},$ we obtain \be \Delta {\bf P}\approx
i{dP_0\over d\zeta}\Omega^{-1}{\bf v}^{\zeta}+c_{s}^{2}{\bf
\rho}_{1}\approx -i\Omega^{-1}\left(c_{s}^{2}\rho {{\partial
}\over {\partial \zeta}}{\bf v}^{\zeta}+c_{s}^{2}\rho {\bf
A}\zeta\right), \label{e24} \ee where we make use of equation
(\ref{e20}) and use the hydrostatic balance condition (\ref{e2}).

The Lagrangian  pressure  perturbation  must
be equal to the radiation pressure
at the surface corresponding to an optical thickness  expected to be of order
unity. As we discuss in Appendix \ref{A}
 we can assume that the radiation pressure acts only on the upper
surface of the disc ( i.e. where $\zeta \rightarrow +\zeta_{0}$). We set,
accordingly, the surface term corresponding to the lower surface of
the disc to zero, and express the term corresponding to the upper
surface in equation (\ref{e23})
with help of equation (\ref{e24}) through  the radiation pressure term
${\bf F}_{+}$ given by equation (\ref{ea10}) of Appendix \ref{A} to obtain
\be\omega r {\bf W}+{F_{0}\over 3\Sigma \Omega }{d \over d
  r}(i{\bf \xi}^{\zeta}/r-{\bf W})+{GM\over 4\Sigma r^{1/2}}
{{d }\over {d r}}\left({\Sigma H^{2} r^{-3/2}\over \hat{ \omega }}{{d }\over
  {d r}}{\bf W}\right)=0, \label{e26}  \ee
where we neglect the second term of the left hand side of (\ref{e23})
proportional to $\omega $ in order to get (\ref{e26}). This procedure is
justified below,  see Section 6.4 where we derive a similar equation
in a more rigorous way.

Also, let us stress that the
$(\zeta )$-component of the displacement vector is assumed to be
evaluated at the surface of the unperturbed disc where the optical
thickness $\tau $ is of the order of unity, and let us recall that
$F_{0}$ is the radiation momentum flux per unit area
perpendicular to the radial direction, see Pringle 1996.

The  density $\rho_{*}(\tau \approx 1)$ where the external radiation
pressure is applied,  is assumed to be
always much smaller than the mid plane density $\rho_{c}$.

We comment that in the absence of radiation pressure, equation
(\ref{e26}) describes the linear dynamics of a free warped disc.
When $\alpha=0,$ this is wavelike, the waves being non dispersive
and having local speed $\Omega H/2$, see  Nelson \& Papaloizou
1999, Papaloizou \&  Lin 1995, Nelson \& Papaloizou 2000.
Furthermore we comment, without giving details,  that equation
(\ref{e26}) may also be derived following the  approach discussed
in those papers, so confirming the analysis based on the adoption
of a twisted coordinate system  presented here.

The radiation pressure term involves the combination ${\bf
\xi}^{\zeta}/r+i{\bf W}.$ This can be regarded as being the
angular displacement of the surface of the disc. It is composed of
two parts. The first $\xi^{\zeta}/r$ represents the displacement
of the surface of the disc relative to the midplane. The second
term $i{\bf W}$ can be viewed as representing the displacement of
the midplane and is described by the twisted coordinate system.
When the disc behaves like a rigid ring, the contribution of ${\bf
\xi}^{\zeta}$ is small.

\subsection{The small parameter $\lambda =\rho_{*}/\rho_{c}$}\label{parlam}

The small parameter $\lambda =\rho_{*}/\rho_{c}$ should be compared
with other small parameters of the problem in order
to specify the different possible regimes of dynamical evolution of the disc.
In particular, as we show below it is important to compare it with the local
 growth rate, in units of $\Omega,$  associated with any possible instability
 of the disc  that is driven by radiation pressure,
 ${\tilde \omega}_{r}.$

As mentioned in the introduction and elaborated further below, we will see, when
$\lambda > {\tilde \omega}_{r},$
we can neglect ${\bf \xi}^{\zeta}$ in the expression  ${\bf \xi}^{\zeta}/r+i{\bf W}$
entering in (\ref{e26}). Physically this corresponds a situation
when the surfaces of constant density in the disc are
approximately parallel
to each other and remain at an approximately constant distance apart.
 In that case, for the purpose of calculation of
the radiation pressure term, the disc may be considered as being
composed of rigid rings having different inclinations and
precession angles and the analysis undertaken in previous studies
(e.g. Pringle 1996) applies. Neglecting  the displacement ${\bf
\xi}^{\zeta}$ in (\ref{e26}), we see that  this equation gives the
evolution of ${\bf W}$ without the need for additional
considerations. For the case $\alpha > \delta $ the frequency
$\omega $ may be neglected in the expression $\hat \omega =\omega
+ i\alpha \Omega $ and equation is reduced to a form analogous to
that
 obtained by Pringle 1996. In this case  Pringle
1996  indicated that the disc may be unstable in linear theory
provided that the radiation pressure term is sufficiently large.

In the opposite limit  $\delta > \alpha, $ equation (\ref{e26}) describes
 wave-like propagation of twisted disturbances of the disc (see
 Nelson \& Papaloizou 1999, Papaloizou \&  Lin 1995,
Nelson \& Papaloizou 2000) as indicated above, modified by the
presence of the radiation pressure term.

\subsection{The polytropic model and the dynamical equation in a
 'standard' form}
For the polytropic equation of state we can bring equation (\ref{e26})
to a  further reduced form. In this case one can obtain from equation
(\ref{e3})
\be \Sigma =\sqrt {\pi}({\Gamma ({\gamma \over \gamma -1})\over
\Gamma ({\gamma \over \gamma -1}+{1\over 2})})\rho_{c}\zeta_{0}, \quad
H^{2}={(\gamma -1)\over (3\gamma -1)}\zeta_{0}^{2}, \label{e27}\ee
where $\Gamma (x) $ is the gamma function.

Before derivation of the general dynamical equation
let us now temporary assume that the disc parameters are such
that $\lambda > \delta^{2}/\alpha $
and the term proportional to ${\bf \xi}^{\zeta}$ in (\ref{e26}) can be neglected
and obtain an equation analogous to what has been obtained by Pringle
(1996) for the polytropic model. In this case we can use equation (\ref{e27})
and rewrite equation  (\ref{e26}) in the form
\be \omega r {\bf W}-{\Gamma ({\gamma \over \gamma -1}+{1\over 2})
\over 3\sqrt {\pi }\Gamma ({\gamma \over \gamma -1})}
\zeta_{0}\tilde F_{0}{{d }\over {d r}}{\bf W}
+{(\gamma - 1)\over 4(3\gamma -1)}{1\over \zeta_{0}\rho_{c} r^{3/2}}
{{d }\over {d r}}\left({\zeta_{0}^{3} \rho_{c}r^{-3/2}\over \hat{ \omega }}{{d }\over
  {d r}}{\bf W}\right)=0, \label{e28}
\ee
where and introduce
the dimensionless radiation pressure parameter
\be\tilde F_{0}={F_{0}\over \rho_{c}\Omega^{2}\zeta_{0}^{2}} \label{e29}\ee
defined as the ratio of the radiation momentum flux $F_{0}$
to a characteristic gas pressure in the mid plane of the disc. In
general, the condition $\tilde F_{0} \ll 1$ should be fulfilled for a
simple description of the evolution of a twisted disc to be valid. Let
us note that when $\gamma =5/3$ the numerical factor in the second term
on the left hand side of (\ref{e28}) is equal to ${8\over 9\pi}$ and
the numerical factor in the last term is equal to ${1\over 24}$.

Equation (\ref{e28}) describes the 'standard' dynamics of a twisted
disc immersed in a radiation field.

\section{The self-consistent approach}\label{sec6}
Now let us take into account the  contribution of all the  terms in  equation (\ref{e26}).
 In order to do that we need   to relate the term in this equation
 involving the $(\zeta)$ component of the Lagrangian displacement vector
to ${\bf W}.$  Also, the term
$2\Omega \omega \int^{+\zeta}_{-\zeta}\rho {v}^{\zeta}d\zeta $ was
discarded without justification  in order to go from equation (\ref{e23}) to
equation (\ref{e26}).  We  provide a formal
justification for this below.
In order to find the vertical component of
velocity, and accordingly, the vertical displacement we  solve
equation (\ref{e22}) considering the right hand side of this equation
to be a source term.  From the appropriate solution of this equation, we
 show that the dynamical equation has indeed the form of
equation (\ref{e26})  to leading  order in our small
parameters $\alpha $ and $1/(\Omega t_{ev})$ and we  also
 specify the form of the vertical displacement at the surface where
the external pressure applies.

\subsection{Solution of the vertical problem for the polytropic model }\label{poly}

In order to find a solution of (\ref{e22}) it is convenient to
separate the vertical velocity ${\bf v}^{\zeta}$ into two parts  such that
${\bf v}^{\zeta}= {\bf v}^{\zeta}_{0}+{\bf v}_{1}^{\zeta}.$

Here the first
part of the decomposition,  ${\bf v}^{\zeta}_{0},$ is a particular solution of (\ref{e22}) with
boundary conditions at the disc surfaces which apply when they are free
and which are regular in the limit of vanishing density.
They are such that in that limit, the Lagrangian
perturbation of the pressure  given by
equation (\ref{e24}) should  approach zero at the disc  surfaces
$\zeta=\pm \zeta_{0}.$

The second part of the decomposition, ${\bf v}_{1}^{\zeta},$ is a solution
of the homogeneous part of equation (\ref{e22}):
\be {{\partial }\over {\partial \zeta}} \left(c_{s}^{2}\rho
{{\partial }\over {\partial \zeta}}{\bf v}_{1}^{\zeta}\right) +2\rho \Omega \omega
{\bf v}_{1}^{\zeta}=0. \label{e30}\ee
This solution will be used in order to satisfy the boundary conditions
\be
\Delta {\bf P}(\rho =\rho_{*},  \zeta  \rightarrow +\zeta_{0})
= {\bf F}_{+}(\rho =\rho_{*}, \zeta \rightarrow +\zeta_{0}), \quad
\Delta {\bf P}(\rho =\rho_{*},  \zeta  \rightarrow -\zeta_{0}) =0,\quad  \label{e30a}\ee
where the Lagrangian perturbation of pressure is given  by equation
(\ref{e24}). We comment that  expressions
such as $\zeta \rightarrow \pm \zeta_{0}$ are used below to apply
to the surface where $\rho =\rho_*,$ the latter being the small density
where the external pressure is applied, rather than at  strictly zero density.

Firstly we derive an expression for the quantity
${v}^{\zeta}_{0}$. To  do this we transform the right hand side of
equation (\ref{e22}) with help of equations (\ref{e3}) and (\ref{e4})
to the form
\be {\bf S}\equiv -2\rho \Omega^{2}\omega r{\bf W}-
\rho \Omega^{2} {\bf A}\zeta^{2}-
{{\partial }\over {\partial \zeta}}(c_{s}^{2}\rho {\bf A}\zeta)=-\rho
\Omega^{2}(\omega r {\bf W}+(\gamma -1){GM\over 4r^{1/2}}{\bf B}),
 \label{e31}\ee
where
\be{\bf B}=\zeta_{0}^{2}\rho_{c}^{-1}{d \over d
  r}\left(\rho_{c}{\bf N}\right)
-3x^{2}\zeta_{0}^{{2\gamma\over \gamma -1}}\rho_{c}^{-1}
{d \over d r}\left(\zeta_{0}^{-{2\over \gamma -1}}\rho_{c}{\bf N}\right),
 \label{e32}\ee
$x=\zeta/\zeta_{0}$ and ${\bf N}=r^{-3/2}(d{\bf W}/dr)/\hat
\omega$. Note that we do not assume that the disc opening angle is
constant when deriving the relations (\ref{e31}) and
(\ref{e32}) and these are valid for any dependence of $\zeta_{0}$ on $r.$

The calculation of ${\bf v}^{\zeta}_{0}$ is greatly facilitated
by noting that the source term  ${\bf S}/\rho$
has a quadratic dependence on $x.$
Is is also easy to see by direct substitution
that  a solution for the vertical velocity
${\bf v}^{\zeta}_{0}$  with   the same quadratic dependence on $x$
can be readily found in the form:

\be {\bf v}^{\zeta}_{0}={\bf b}_{1}+{\bf b}_{2}x^{2}, \label{e33} \ee
where expressions for the quantities ${\bf b}_{1}$ and ${\bf b}_{2}$
directly follow from equation (\ref{e22}) and the form of the source
term (\ref{e31})
\be {\bf b}_{1}=-r\Omega {\bf W}-{(\gamma -1)\over (3\gamma -1)}
{GM\over 4r^{1/2}}{\tilde \omega}^{-1}\zeta_{0}^{-1}\rho_{c}^{-1}
{d \over d r}\left(\zeta_{0}^{3}\rho_{c}{\bf N}\right ), \label{e34}\ee
and
\be {\bf b}_{2}=-{3\over 4}{(\gamma -1)\over (3\gamma -1)}
{GM\over r^{1/2}}\zeta_{0}^{{2\gamma \over \gamma -1}}
{d \over d r}\left(\zeta_{0}^{-{2\over \gamma -1}}\rho_{c}{\bf N}\right),
  \label{e35}\ee
where we introduce the dimensionless frequency $\tilde \omega = \omega
/\Omega $ and assume that it is small: $|\tilde \omega |\ll 1$.
In this limit it is easy to see that the quantity ${\bf b}_{2}$ is
$\sim \tilde \omega $ times smaller than ${\bf b}_{1}$. It is, therefore,
neglected later on, and we assume that
\be {\bf v}^{\zeta}_{0}\approx {\bf b}_{1} \label{e33a} \ee
in our future analysis.

\subsection{The solution of the homogeneous problem}
Now let us  find the homogeneous  component ${\bf v}_1^{\zeta}$.  It turns
out that the solutions of  equation (\ref{e30})  take on a
very simple form  when
$\gamma =5/3.$  In that case the solutions can be expressed in terms of
elementary functions. Therefore, we  shall specialise to  this case in the main
text of the paper from now on. However, the final description of the system we obtain
has wider applicability (see below).
The form of the solutions of the homogeneous problem
corresponding to  general values of $\gamma ,$ which can be used
to construct the full solution in that case is relegated  to Appendix \ref{B}.

When
$\gamma =5/3$ it is convenient to introduce new variables $\theta $ and
${\bf Y}$ according to the  prescription
\be \zeta =\zeta_{0}\sin \theta, \quad {\bf v}^{\zeta}={\bf Y}/\cos^{3}
\theta. \label{e36}\ee
It is obvious from equation (\ref{e36}) that  $\theta
=0$ and $\theta =\pi/2$
correspond to the disc midplane $\zeta=0$ and to the
free surface  $\zeta =\zeta_{0}$, respectively. It follows
from equation (\ref{e3}) that the density is proportional to $\cos^{3}
\theta $ and  accordingly the variable $\bf Y$ is proportional to
$\zeta$-component of the gas momentum per unit volume.

In terms of the  variables defined through  (\ref{e36}), equation  (\ref{e30})
takes the form
\be {d^{2} \over d\theta^{2}}{\bf Y}+2\tan \theta  {d\over d\theta
}{\bf Y} +3(1+2\tilde \omega ){\bf Y}=0. \label{e37}\ee
It can be verified by direct substitution that the two independent
solutions of (\ref{e37}) are
\be {\bf Y}_{1}=(\kappa +1)\cos (\kappa -1)\theta +(\kappa -1)\cos
(\kappa +1)\theta , \quad  {\bf Y}_{2}=(\kappa +1)\sin (\kappa -1)\theta
+(\kappa -1)\sin (\kappa +1)\theta , \label{e38}\ee
where $\kappa =\sqrt {4+6\tilde \omega }$. Note that the solutions
(\ref {e38}) are valid for any value of $\tilde \omega $. For our
purposes, however, only the case $|\tilde \omega |\ll 1$ is significant
and we accordingly set $\kappa = 2 +{3\over 2}\tilde \omega $ and
take into account only zeroth and first order terms when
summing series in ascending powers of  $\omega$\footnote{Contrary to
the inhomogeneous part ${\bf v}^{\zeta}_{0}$ the next order terms in
$\tilde \omega $ in the  homogeneous part
diverge near the disc surface. They can
be of order of or larger than the leading terms and must, therefore,
be retained.}.

Thus the general solution of equation (\ref{e22}) can be expressed
as a linear combination of these independent solutions
in the form
\be { v}^{\zeta}=(C_{1}{\bf Y}_{1}+C_{2}{\bf Y}_{2})/\cos^{3}\theta+{
v}^{\zeta}_{0}, \label{e39}\ee
where ${ v}^{\zeta}_{0}$ is given by equation (\ref{e33}) with
$\gamma =5/3$ and the arbitrary
constants $C_{1}$ and $C_{2}$ can be chosen to satisfy the boundary
conditions (\ref{e30a}).

\subsection{A specific gauge freedom associated with the twisted coordinate system}\label{gaugef}

Following the simple approach
of section \ref{simpa}, a dynamical equation for
the quantity ${\bf W}$ can be obtained from equation (\ref{e22}) by
integration over $\zeta $  provided that certain terms are discarded. However,
in a more exact approach, it
is easy to see that this equation cannot be used  for
determining the evolution of  ${\bf W}$
without invoking some additional considerations.
As we have mentioned above this  situation  comes about because there is some degree of
arbitrariness  associated with the specification of the
twisted coordinate system used in our study.

The resulting gauge freedom may be used in order to put constraints on some dynamical
variables by choosing the most appropriate gauge
from physical point of view. In our case it seems reasonable to look
for a gauge where the vertical component of velocity ${\bf v}^{\zeta}$
has a small absolute value. This could be specified by imposing the
requirement that
\be { v}^{\zeta}(\zeta =0)=0. \label{e40}\ee
As we will see below the condition (\ref{e40}) together with equation
(\ref{e22}) fully specify all dynamical variables and allow us to
obtain a single equation determining the dynamical evolution of  ${\bf W}.$
\footnote{Note that
the condition (\ref {e40}) is not unique. Another reasonable
condition would be eg. the requirement that
$\int^{+\zeta_{0}}_{-\zeta_{0}}d\zeta \rho {\xi}_{\zeta}=0.$ We expect,
however, that any reasonable condition used to fix the gauge would give the
same dynamical equation to leading order in the small
parameter $1/( \Omega t_{ev}).$}

\subsection{Derivation of the dynamical equation}\label{Dyneq2}

In order to find the dynamical equation we  calculate the
coefficients $C_{1} $ and $C_{2}$ entering (\ref {e39}) with help of
equations (\ref {e24}), (\ref {e30a}) and (\ref {ea10}). Then
application of the gauge condition (\ref {e40}) will give us the dynamical
equation.

We note that only the homogeneous part ${\bf v}^{\zeta}_{1}$
of the vertical velocity can lead to a non vanishing
Lagrangian  pressure perturbation
in the limit of zero density
when $\zeta \rightarrow \pm \zeta_{0}.$
Thus this component of the solution dominates near the surface. Using equations
(\ref {e24}), (\ref {e30a}) and (\ref {ea10}), in the limit of small surface
density, $\rho_*,$  we get
\be{\partial \over \partial \zeta}{\bf v}^{\zeta}_{1}(\zeta \rightarrow
+\zeta_{0})=-{2\over 3}{F_{0}\over c_{s}^{2}\rho_{*}}\Omega r{d\over d
  r}\left({{\bf v}^{\zeta}_{1}\over \Omega r}+{\bf W}\right ), \quad
{\partial \over \partial \zeta}{\bf v}^{\zeta}_{1}(\zeta \rightarrow
-\zeta_{0}) =0,
\label{e41}\ee
where we assume that all quantities are evaluated at $\rho=\rho_{*}.$

Differentiating expression (\ref {e39}), using  (\ref {e38}) remembering
that $\zeta =\zeta_{0}\sin \theta $ and considering the limit $\zeta \rightarrow
\pm \zeta_{0}$ and accordingly $\theta \rightarrow \pm \pi/2$ we approximately have
\be {\partial \over \partial \zeta}{\bf v}^{\zeta}_{1}(\zeta \rightarrow
\pm \zeta_{0}) \approx {3\over \zeta_{0}\cos^{5} \theta }(\mp {3\over 2}\pi
  \tilde \omega C_{1}+2C_{2}). \label{e42}\ee
The second condition in (\ref {e41}) tells that $C_{2}=-{3\over 4}\pi
\tilde \omega C_{1}$. Using this fact we obtain from equations (\ref{e33a}),
(\ref {e38}),  (\ref {e39}) and (\ref {e42}) in the same limit
\be  {\partial \over \partial \zeta}{\bf v}^{\zeta}(\zeta \rightarrow
 +\zeta_{0})\approx {\partial \over \partial \zeta}{\bf v}^{\zeta}_{1}(\zeta \rightarrow
 +\zeta_{0}) \approx -{9\pi \tilde \omega \over \zeta_{0}\cos^{5} \theta
 }C_{1}, \label{e43}\ee
and
\be {\bf v}^{\zeta}={\bf v}^{\zeta}_{0}+{\bf v}^{\zeta}_{1}\approx {\bf b}_{1}+C_{1}(4 -{3\pi \tilde \omega \over
\cos^{3}\theta})= {\bf b}_{1}+C_{1}(4-{3\pi \tilde \omega \over \lambda }), \label{e43a}\ee
where we recall that $\lambda \equiv \rho_{*}/\rho_{c}=\cos(\theta )$ such that
the coordinate $\theta $ corresponds to the density level where $\rho = \rho_{*}.$

Using equations (\ref{e3}) and  (\ref{e4}) with $\gamma =5/3$ we can
represent the factor in the front of the radial derivative on the right hand side
of equation (\ref{e41}) in the form
\be {2\over 3}{F_{0}\over c_{s}^{2}\rho_{*}}\Omega r={2\tilde
F_{0}\over \cos^{5} \theta}\Omega r. \label{e44}    \ee
Finally, substituting equations (\ref{e43}) and (\ref{e43a})  in equation (\ref{e41})
and taking into account (\ref{e44}) we obtain
an equation for the quantity $C_{1}$
\be \epsilon \delta^{2}\Omega r^{2}{d\over d r}\left({\tilde \omega C_{1}\over \lambda
  \Omega r}-{1\over 3\pi}({\bf b}_{1}+4C_{1}+{\bf W})\right)+{3\over 2}\tilde \omega C_{1}=0,
\label{e45}\ee
where we introduce an important parameter
\be \epsilon ={r\over \zeta_{0}}\tilde F_{0}={r F_{0}\over
  \rho_{c}\Omega^{2}\zeta^{3}_{0}}. \label{e46}\ee
This parameter determines strength of dynamical effects caused by the
radiation pressure acting on the disc. When $\epsilon > 1$ these effects
may be significant. Physically, the inequality $\epsilon > 1$ means that
the ratio of the momentum flux carried by radiation, $F_{0}$, to the
characteristic mid plane gas pressure in the disc is larger than the
disc opening angle.

Now let us consider the gauge condition (\ref{e40}). From this
condition and the expression (\ref{e33}) it follows that
\be C_{1}{\bf Y}_{1}(\theta =0)=-{\bf b_{1}}, \label{e47}\ee
where we use the variable ${\bf Y}_{1}$ defined in equations
(\ref{e36}) and (\ref{e38}). Taking into account that ${\bf
Y}_{1}(\theta =0)=4$ and, accordingly ${\bf b}_{1}+4C_{1}=0$ in
equation (\ref{e45}) and using the explicit form for the quantity
${\bf b}_{1}$ given in equation (\ref{e34}) with $\gamma =5/3$ we
obtain
\be C_{1}={1\over 4}\left(r\Omega {\bf W}+
{GM\over 24r^{1/2}}{\tilde \omega}^{-1}\zeta_{0}^{-1}\rho_{c}^{-1}
{d \over d r}(\zeta_{0}^{3}\rho_{c}{\bf N})\right). \label{e47a}\ee
Equations (\ref{e45}) and (\ref{e47a}) form a complete system for describing
the disc dynamics. Substituting $C_{1}$ given by equation (\ref{e47a})
into equation (\ref{e45})  we can obtain a single third order equation
for the quantity ${\bf W}$. However, we find it more convenient
to retain the pair of equations obtained directly from  (\ref{e45}) and (\ref{e47a})
to describe the disc dynamics.

\subsection{Description of the dynamical system}\label{Dyneq}

In order to bring this pair of equations  into
a simpler form, we introduce a new dimensionless variable
${\bf x}={4\tilde \omega \over \Omega r}C_{1}$. Using ${\bf x}$
 to replace $C_1$ in  equations (\ref{e45}) and (\ref{e47a}) we obtain
\be {3\over 2}{\bf x} = \epsilon \delta ^{2}r\left(
{4\over 3\pi}{d \over d r}{\bf W}-{d \over d r}(\lambda^{-1}{\bf x})\right)
\equiv {\epsilon \delta ^{2}\rho_c\zeta_0 r\over 2\Sigma}\left(
{d \over d r}{\bf W}-{d \over d r}(2\Sigma(\rho_c\zeta_0\lambda)^{-1}{\bf x})\right), \label{e48}\ee
and
\be {\bf x}-\tilde \omega {\bf W} ={(GM)^{1/2}\over 24\zeta_{0}\rho_{c}}
{d \over d r}\left({\zeta^{3}_{0} \rho_{c} r^{-3/2}\over \hat \omega }{d
  \over d r}{\bf W}\right)
  \equiv  {(GM)^{1/2}\over 4\Sigma }
{{d }\over {d r}}\left({\Sigma H^{2} r^{-3/2}\over \hat{ \omega }}{{d }\over
  {d r}}{\bf W}\right), \label{e49}\ee
where we have used the explicit expression for the quantity ${\bf N}$ defined through
equation (\ref{e32}) and
recall that $\hat \omega =\omega + i\alpha \Omega =\Omega (\tilde
\omega +i\alpha)$.

It is easy to see that equation (\ref{e26}) follows from
 form  equations (\ref{e48}) and (\ref{e49}) provided that the $(\zeta)$-component of the
Lagrangian displacement vector in  equation (\ref{e26})  is replaced by
\be {\bf \xi}^{\zeta}=-{3\pi i\over 4}\lambda^{-1}r{\bf x}. \label{e50}\ee

In order to be able to neglect this term in equation (\ref{e48}),
we require that $|{\bf x}| \ll |\lambda {\bf W}|.$
When this is satisfied, ${\bf x}$ can be determined after neglecting the
first term in this equation, and assuming that ${\bf W}$ varies
on a length scale comparable to $r,$ with the result that
${\bf x} \sim \epsilon \delta^2.$ Thus a description assuming that the
vertical response of the disc is rigid
can be recovered only when the disc parameters are such that
$\lambda \gg \epsilon\delta^2.$
This is precisely the condition we obtained from a simple argument
given in the introduction.

In the opposite limit the
contribution proportional to  ${d \over d r}(\lambda^{-1}{\bf x})$
should not be neglected in equation (\ref{e48}) when the effects
of the radiation pressure are significant.
Then as $\lambda \rightarrow 0,$ this equation gives
\be {\bf x}={4 \over 3\pi }\lambda{\bf W}.\ee
Using  the above to eliminate ${\bf x}$ in equation (\ref{e49})
we obtain an equation describing stable radially propagating
disturbances which attain a frequency given by
${\tilde \omega} ={4 \over 3\pi }\lambda$ in the limit of zero radial wave number
(see also the discussion below).

We comment here that although we used a polytropic model to derive
equations (\ref{e48}) and (\ref{e49}) and continue with the immediate discussion,
the indicated equivalent  description of
the dynamical system, which uses $\Sigma$ and $H$  with the numerical factor $3\pi/4$ replaced by
$2\Sigma/(\rho_c\zeta_0),$ contains no explicit dependence on the local
details of the model, other than through the product of the height at which
the external pressure is applied and the value of the density there,
$\Sigma$ and $H.$
This shows that our formalism
has a more general applicability and we have verified this
to be the case by obtaining the equivalent representation, quite generally, using
an alternative  analysis that does not use twisted coordinate systems,
see e.g.  Nelson \& Papaloizou 1999,
Papaloizou \&  Lin 1995
Nelson \& Papaloizou 2000.

\section{Qualitative analysis of the dynamical equation}\label{qualan}

\subsection{Local analysis}

In general, equations (\ref{e48}) and (\ref{e49}) should be solved
numerically, for a specified  disc model. However, this is beyond
the scope of the present paper. Here we follow Pringle 1996 and
adopt the usual local approximation scheme in which it is assumed
that the radial dependence of  all variables is $\propto e^{ikr}$
and that the radial wave-number $k$ is such that $|kr| \gg 1.$
Note however, that although details must depend on global issues and
boundary conditions so that one has to proceed with caution, we
may also expect that this analysis gives reasonable order of
magnitude estimates even when  $|kr| \sim 1$.

In addition, for simplicity we  shall also continue to consider
 the polytropic model as the
essential physics is contained therein,
while bearing mind that
results, when expressed only in terms of the local
 disc model parameters, $\Sigma, H,$ and $\rho_*\zeta_0,$ 
 can be straightforwardly applied to 
other models on the basis of the discussion
given in section \ref{Dyneq} above.

Setting ${\bf x}, {\bf W} \propto  e^{ikr}$ in equations
(\ref{e48}) and (\ref{e49}) we obtain
\be \left(1+{2i\over 3}{\epsilon \delta ^{2}\over \lambda }\tilde k\right){\bf
x}={8i\over 9\pi}\epsilon \delta^{2}\tilde k {\bf W}, \label{e51}\ee
 and
\be {\bf x}=\left(\tilde \omega -{\delta^{2}\over 24}
{\tilde k^{2}\over (\tilde \omega + i\alpha )}\right){\bf W} \label{e52}\ee
respectively, where $\tilde k =kr$ is the dimensionless
wave number. One can use equations (\ref{e51}) and (\ref{e52})
to obtain dispersion relation $\tilde \omega = \tilde \omega (\tilde
k).$ However, as we have to consider several
limiting cases for which there is qualitatively different dynamics,
the general form has limited usefulness.

\subsubsection{Dynamics of a  twisted disc without external radiation pressure:
Viscous and wave regimes}

At first let us recall the basic dynamical properties of the standard twisted
disc where  radiation pressure effects are absent. In this case the
dispersion relation can be easily obtained from equation (\ref{e52})
together with  the condition ${\bf
x}=0.$ We then obtain
\be \tilde \omega ={1\over 2}(-i\alpha \pm \sqrt { \delta^{2}\tilde
k^{2}/ 6-\alpha^{2}}). \label{e53}\ee
The behaviour of the disc is qualitatively different in the two limiting
cases of sufficiently large and sufficiently small  $\alpha $ parameter
(PP, Papaloizou $\&$ Lin 1995). For the case of a large $\alpha $
the 'slow' mode corresponding to the the positive imaginary part in (\ref{e53}) determines
the disc dynamics and the dispersions relation can be approximately
written as
\be  \tilde \omega \approx -i\tilde \omega_{\nu}, \quad
\tilde \omega_{\nu}={\delta^{2}\tilde k^{2}\over 24 \alpha}. \label{e54}\ee
Remembering that according to our convention all quantities are
proportional to $e^{-i\omega t}\approx e^{-\omega_{\nu} t}$ we
conclude that in this case, twisted perturbations of the disc decay. In the
 limit of small $\alpha$  it is easy to see that  twisted perturbations
propagate over the disc with a speed of the order of a typical
sound speed with little dissipation  and we have
\be \tilde \omega \approx \pm \tilde \omega_{s} -i\alpha/2, \quad
\tilde \omega_{s}={\delta
|\tilde k|\over 2\sqrt{6}}. \label{e55}\ee
From the condition $\tilde \omega_{\nu} < \tilde \omega_{s}$ one can see that the
'viscous' regime of evolution of the disc is realised when
\be \alpha > {\delta |\tilde k|\over 2\sqrt{6}}. \label{e56}\ee
For the case of a 'global' perturbation with $|\tilde k| \sim 1$ the
above condition is approximately reduced to requirement that the viscosity
parameter $\alpha $ is larger than the disc opening angle $\delta $.

\subsubsection{The effects of  radiation pressure for relatively large $\alpha $}\label{7.1.2}

Now let us consider effects induced by external radiation pressure.
At first let us assume that the condition (\ref{e56}) is fulfilled.
In this case we  have from equation (\ref{e52})
that approximately
\be {\bf x}\approx (\tilde \omega +i\tilde \omega_{\nu}){\bf W}. \label{e57}\ee
Substituting equation (\ref{e57}) in (\ref{e51}) we obtain
\be \tilde \omega ={1\over (1+{4\over 9}\left({\tilde \omega_{r}\over
\lambda}\right)^{2})}\left({16\over 27\pi\lambda}\tilde \omega_{r}^{2} +i{8\over
9\pi}\tilde \omega_{r}\right) -i\tilde \omega_{\nu}, \label{e58}  \ee
where we now define the  time
scale characterising influence of the radiation pressure,
${\tilde \omega}_{r}^{-1},$  such that
\be \tilde \omega_{r}=\epsilon \delta^{2}\tilde k. \label{e59}\ee
The dimensionless time scale ${\tilde \omega}_{r}^{-1}$ is that associated with the
radiation pressure driven  instability
considered by Pringle 1996.

It is clear from equation (\ref{e58}) that the character of  the dynamical evolution
in fact depends on a comparison of  $\tilde \omega_{r} $ to $\lambda = \rho_*/\rho_c. $
As $\tilde \omega_{r} = \epsilon \delta^2 {\tilde k},$ this leads directly
to the condition $\epsilon \delta^2 \ll \lambda$ being required for the assumption
that the vertical response of the disc is that of a rigid body to be valid
for disturbances with radial scale comparable to $r.$ That condition was also
derived on general  grounds in section \ref{Dyneq} and from simple 
arguments in the introduction and is explored further below.

First let us assume the  case  $\tilde \omega_{r} \ll \lambda$.
In this limit the factor in the front of the first brackets in
(\ref{e58}) is equal to one and because by necessity $\lambda < 1$, the first term
in  the brackets can be neglected, and therefore $\tilde \omega $
is  purely imaginary such that
\be \tilde \omega = i({8\over 9\pi}\tilde \omega_{r}-\tilde
\omega_{\nu}). \label{e60}\ee
This gives either growth or decay of the perturbation depending on
whether the sign of the expression in the brackets is positive or
negative. Since $\tilde \omega_{r}$ is proportional to $\tilde k$ it can be,
in principal, positive when $\tilde k > 0$.
Thus one can infer  instability  of the
disc as has been done by  Pringle 1996. The quantity $\lambda$ drops out
and the disc behaves as a collection of rigid rings.
 The corresponding stability criterion can be obtained from the
condition that the absolute value of the first term in the brackets is
larger than the second term and therefore
\be \epsilon > \epsilon_{crit}={3\pi\over 64}{|\tilde k |\over \alpha}. \label{e61}\ee
Note that the numerical factor, $\sim 0.1,$ in the expression for
$\epsilon_{crit}$ depends on the vertical structure of the disc. It
seems reasonable to assume, however, that it is always smaller than
one for any possible vertical distribution of pressure and density.

Let let us now  consider the  limit $\tilde \omega_{r} \gg
\lambda$. In this limit  the character of the disc evolution changes drastically.
This is essentially because the vertical motion can no longer
be treated as uniform as would occur for a series of rigid rings.
 Instead of potential  growth of disc
perturbations, we  find mainly decaying oscillatory dynamics. The
corresponding dispersion relation reads
\be \tilde \omega ={4\over 3\pi}\lambda +i({2\over
  \pi}{\lambda^{2}\over \tilde \omega_{r}} -\tilde \omega_{\nu}). \label{e62}\ee

As seen from (\ref{e62}) the real part of the frequency does not
depend on the magnitude of the radiation pressure. The imaginary part  can be either
negative or positive. The values of the terms in the brackets,
being always negative in the limit of very small $\lambda$ or
 very large $\tilde \omega_{r}$
and  thus very large $F_{0},$ lead  to damping in those limits.
In any event we might also expect that the
oscillations in the disc inclination could result in strong
self-shielding of the disc which can inhibit the growth of the inclination
angle.

\subsubsection{The  effects of radiation pressure for small $\alpha $}

Let us now consider the opposite case of a small viscosity parameter
$\alpha $ such that the inequality (\ref{e56}) is not satisfied. In
this case, provided that the value of radiation flux is smaller than
the typical mid plane pressure of the gas, a simple analysis shows that the effect of
 radiation pressure is to  give a correction to the dispersion relation
(\ref{e55}).

We begin by pointing out that equation (\ref{e52}) in
this limit  can be brought to the form
\be {\bf x}\approx {1\over \tilde \omega}(\tilde \omega^{2}-{\delta^{2}\tilde
k^{2}\over 24}+i{\alpha \delta^{2} \tilde k^{2}\over 24\tilde
\omega}){\bf W}, \label{e69}\ee
where the last term in the bracket is a small correction.  After
substitution of (\ref{e69}) in equation (\ref{e51}) we can look for
solution for the dispersion relation  in the form;
$\tilde \omega =\pm \omega_{s}-i\alpha +\tilde \omega_{1}$.
It is easy to see that the correction $\tilde \omega_{1}$ has the form
\be \tilde \omega_{1}={4\over 9\pi}{\tilde \omega_{r}\over (i+{2\over
3} \tilde \omega_r / \lambda)}. \label{e70} \ee

It is clear from (\ref{e70}) than we always have $|\tilde \omega_{1}|
< {4\over 9\pi}\tilde \omega_{r}$.  In addition we note that
\be |\tilde \omega_{r}/\tilde \omega_{s}| \sim \epsilon
\delta. \label{e71}\ee

As follows from the definition of the parameter $\epsilon $ (see
equation (\ref{e46})), the right hand side of (\ref{e71}) is smaller
than unity provided that the radiation flux is smaller than the disc
midplane pressure. Assuming that this is valid the disc dynamics, in the
case of the small $\alpha,$ is mainly determined by propagation of
waves with a speed of the order of the sound speed as in the case
without radiation pressure. As in the case of  very small
$\lambda $ considered above, we again get damping in that limit.
 It also seems reasonable to suppose that
in this situation, fast oscillations of the inclination angle would lead
to a strong self-shielding of the disc thus inhibiting possible
instabilities.

\subsection{Physical character of the disc dynamics in the regime
where $\lambda \ll \tilde \omega_{r}$}\label{smalllam}

As  discussed above,  when $\lambda \ll \tilde \omega_{r}$ the
character of the disc dynamics  differs drastically from that of a
set of rigid rings. The disc then tends to oscillate at the
frequency $\tilde \omega \approx {4\over 3\pi}\lambda$, see
equation (\ref{e62}) and the discussion in section \ref{Dyneq}.
Since this regime has not been discussed elsewhere we briefly
describe its main features.

Assuming  the disc to be in this regime, as indicated in section
\ref{Dyneq}, we can neglect unity in
the brackets on the right hand side of equation (\ref{e51}) and obtain
a simple relation between ${\bf x}$ and ${\bf W}$ such that
\be {\bf x}\approx {4\lambda \over 3\pi}{\bf W}. \label{e72}\ee
Now we use equation (\ref{e72}) and  equation (\ref{e50}) together with
 equation (\ref{ea10}) of Appendix \ref {A} to see that the radiation
pressure term ${\bf F}_{+} \rightarrow 0$ when  $\lambda / \tilde
\omega_{r} \rightarrow 0$.

Therefore, in this limit the disc tends to evolve in such a
way that its surface becomes almost parallel to the initial or radial direction thus
strongly decreasing the amount of any increase in the amount of intercepted radiation
as a result of the perturbation. The disc gas does not cross this
surface and therefore, in this limit, the radiation pressure effectively provides
an impenetrable wall at the density $\rho_{*}$ thus prohibiting the gas motion
in vertical direction. However, when there is no influence
of the radiation pressure on the lower disc boundary, as mentioned already in section \ref{Dyneq},
stable radially propagating warp like disturbances can still still exist.

\subsection{Maximal potential growth/decay rate}\label{7.3}

An interesting consequence of the effects described above is that,
when equation (\ref{e58}) is considered, there is an extremum of the
imaginary part of $\tilde \omega $, $\tilde \omega_{I}$ as a function
of $\tilde \omega_{r}$. The extremum is attained when
$\tilde \omega_{r} = \tilde \omega_{r} ^{e}= \pm {3\over 2}\lambda $ and the corresponding
value of $\tilde \omega_{I}$ is equal to
\be \tilde \omega_{I}^{e}=\pm {2\over 3\pi}\lambda -\tilde
\omega_{\nu}. \label{e73}\ee
Thus  for any value of the external radiation pressure and corresponding
positive/negative  $\omega_{r}$ corresponding to the instability/decay
of the disc perturbations, the growth/decay rate cannot exceed the value
given by equation (\ref{e73}) with the signs $(+)$ and $(-)$
corresponding to  instability and decay, respectively.

When the effect of the external  radiation pressure is significant,
the second term on the right hand side of (\ref{e73}) can be neglected
and we have a very simple expression for the maximal absolute value
of the potential growth/decay rate
\be |\tilde \omega_{I}^{e}|={2\over 3\pi}\lambda \label{e74}.\ee

\subsection{ Conditions on disc model parameters for potential instability}\label{condi}

As indicated above, the disc dynamics
 in the presence of the radiation field
 is more complex in the case of a relatively large $\alpha $ for which
inequality  (\ref{e56}) is fulfilled. Then, different dynamical regimes
 are possible depending on the parameters of
the unperturbed disc  model as well as the magnitude of the external
radiation flux. Here we  estimate the
characteristic parameter values that separate the different
dynamical regimes assuming that inequality (\ref{e56}) is valid.

According a  the linear theory of twisted disc perturbations
that makes use of a local WKB analysis,
instability associated with the effect of radiation
pressure appears to be unaffected by the disc density $\rho_*$
at the location  it is applied when
$|\tilde \omega_{r}| \ll {3\over 2}\lambda$.
When this condition  is not satisfied, the disc dynamics is strongly affected by
effects associated with the non-rigid body  character of the response of the  disc
vertical structure and the potential instability may be either
inhibited or absent.

The above inequality leads to an upper limit for
the value $\epsilon $ and, accordingly, on the value of the radiation
momentum flux $F_{0}$
\be \epsilon < \epsilon_{*}={3\over 2}{\lambda \over \delta^{2}|\tilde
  k|}. \label{e63}\ee
On the other hand the quantity $\epsilon $ should be larger than
$\epsilon_{crit}$ given by equation (\ref{e61}). Therefore,
instability seems to be possible only for a certain range of values
of the radiation momentum flux. From the condition $\epsilon_{crit} <
\epsilon_{*}$ we find a condition on the disc parameters for the
instability to be possible in the form
\be {\lambda \alpha \over \delta^{2}} > {\pi \over 32}|k|\approx
0.1|k|. \label{e64}\ee
Assuming that this estimate is approximately valid even when $|\tilde
k| \sim 1$ we obtain a lower bound for the value of the density ratio
$\lambda $ from (\ref{e64})
\be \lambda \equiv {\rho_{*} \over \rho_{c}} > \lambda_{crit}=
{10^{-4}\delta^{2}_{(-2)}\over \alpha_{(-1)}}, \label{e65}\ee
where $\delta_{(-2)}=\delta/10^{-2}$ and $\alpha_{(-1)}=\alpha/10^{-1}$.
When the density ratio is very small and such that $\lambda < \lambda_{crit},$ the
simple theory indicates that the
disc is likely to be stable for the whole range of possible values of
the radiation flux $F_{0}.$

\subsection{Some conditions for non-linearity}\label{7.5}
We can obtain a value $F_0 = F_{*}$ of the radiation momentum flux corresponding to
the characteristic value $\epsilon_{*}$ separating two possible regimes of
evolution of the disc in the presence of  radiation pressure.  By making use of
 equation (\ref{e46}) we find
\be F_{*}\approx {\lambda \over \delta }P_{c}, \label{e66}\ee
where $P_{c}=\rho_{c}\Omega^{2}\xi_{0}^{2}$ estimates the  gas
pressure in the disc mid plane. As  mentioned above, in a
situation where flaring of the disc is present, the magnitude of the
radiation momentum flux should be smaller than  $P_{c}$ in order
for our simple theory to be approximately correct.
This sets an upper
bound for the value of $\lambda ,$ such that $\lambda < \delta .$
Assuming that the radiation momentum flux, $F_{0},$ is near to
 $F_{*},$ the expression (\ref{e66}) leads to a stringent constraint on the
value of the inclination angle $\beta .$ In order for the
linear theory to be valid, a characteristic  value of the radiation pressure in the
disc $F\sim F_{0}r(d\beta/dr) $  should be smaller than the value of the gas
pressure at the surface where the radiation pressure is applied,
$P_{*}\equiv P(\rho_{*})\approx P_{c}\lambda^{5/3}$ for the polytropic model
with $\gamma =5/3.$

From the
condition $F < P_{*},$ assuming that $\beta$ changes on scale comparable
to the radius, we obtain a condition for linear theory to be valid
in the form
\be \beta < \beta_{*}=\lambda^{2/3}\delta \ll 1. \label{e67}\ee
Note, however, that this constraint has a significant dependence  on the
vertical structure of the disc. Thus, for an isothermal disc where
$P \propto \rho $ the small parameter $\lambda^{2/3}$ in
(\ref{e67}) would be absent.

It is also interesting to note that for a polytropic disc, similar
considerations lead to a strong constraint on the value of $\beta
$ when the radiation flux $F_{0}$ is of the order of $F_{crit}$
corresponding to onset of a radiation pressure induced instability
in the disc. In this case we can repeat the above argument
 but using the value $\epsilon_{crit}$ given
by equation (\ref{e61}) with $\tilde k =0$ instead of
$\epsilon_{*}$.

Proceeding this we find that linear theory would be valid only if
\be \beta < \beta_{crit}=10\lambda^{5/3}\alpha/\delta. \label{e68}\ee
Substituting typical values $\delta=10^{-2}$ and $\alpha
=10^{-1}$ and assuming that $\lambda \sim 10^{-3}$ we get
$\beta_{crit}=10^{-3}.$  Thus, for unperturbed disc models with strong
density and temperature drops toward the surface boundary of the disc,
the linear theory of perturbations is, strictly speaking, valid only
for quite small values of the inclination angle.

\subsection{The one dimensional vertically stratified slab analogue}\label{slaban}
As we discuss in more detail in the next section
there is an analogy between the motion in this regime and that
occurring in the simple hydrodynamic
problem of free one-dimensional oscillations of a vertically  stratified column of
gas, with equation of state (\ref{e1}),
gravitational acceleration $g=-\Omega^{2}\zeta,$ lower free boundary,
 and an upper rigid boundary at some density
$\rho_{*}\ll \rho_{0}$.
The unperturbed distributions of pressure and density are
given by equation (\ref{e3}). Assuming that perturbed quantities
depend on time through a factor  $\propto e^{-i(\Omega +\omega)t}$ the free oscillations
of the gas column are described by equation (\ref{e30}) and therefore
expressions (\ref{e38}) and (\ref{e39}) with the response ${\bf {v}}^{\zeta}_{0}=0$
provide the solutions of the free problem for a polytrope with $\gamma =5/3.$
As above we shall focus on that case below.

If we assume that such a column is rotating around a  central body
with  angular frequency $\Omega$ and $m=1,$  the frequency of oscillations
observed in  a non rotating  frame will be equal to $\omega.$  As we will see
later the smallest eigen value $\omega,$ for the
normal mode oscillations of this problem, is precisely
equal to ${4\lambda \over 3\pi}\Omega$ - the frequency of the disc
oscillations, when the radial wavelength is very long,
 and it is in the regime $\lambda \ll \tilde \omega_{r}$.

\section{Linear and non linear dynamics of the vertically stratified slab}\label{8}

When the condition $\tilde \omega_{r} \gg\lambda$ is fulfilled,
 internal non uniform vertical motions
determine the disc dynamics.  As we have
mentioned in the previous section, the dynamics of the disc
in the long radial wavelength limit is quite
similar to the dynamics of a perturbed vertically stratified
 one-dimensional gas column with
equation of state given by (\ref{e1}) and
gravitational acceleration $g=-\Omega^{2}\zeta.$

Contrary to the disc case, the one dimensional problem can be
easily studied, both analytically and numerically. The results  can be used  to check the validity
of some of our assumptions
leading to conclusions about the disc dynamics. They can also provide
some understanding of the disc dynamics in the non-linear regime
when the external pressure perturbation is of the order of disc pressure at the
density level where the external forcing is applied.

It turns out that the
analogy between the disc problem and the one dimensional column problem is practically
exact in  linear perturbation theory, provided that we consider
perturbations of the column subject to the upper boundary condition
\be \Delta P\equiv P-P_{0}=a({P_{e}\over \Omega \zeta_{0}})v,
\label{e75}\ee
where $P_{0}$ is the equilibrium distribution of pressure given by
equation (\ref{e3}), $v$ is the gas velocity in the vertical direction,
 $P_{e}$ is the externally applied pressure and
$a$ is a dimensionless parameter. But note that if unphysical
behaviour is to be avoided with this boundary condition, we
require $a < 0,$ see below. All upper boundary  quantities are
assumed to be evaluated at some coordinate $\zeta_{*}$
corresponding to the density $\rho_{*}\equiv \rho(\zeta
=\zeta_{*}) \ll \rho_{c}.$
 Note that in the limit $a\rightarrow \infty ,$
we must have $v(\zeta =\zeta_{*})=0$ in order to have a finite
 pressure at the boundary. Then the condition (\ref{e75})  reduces
to  a rigid wall condition. The lower boundary remains free.

We go on to consider the
dynamics of the vertically stratified column in detail,
both analytically and numerically.
As in the disc case we set
$\gamma =5/3.$

\subsection{The eigenvalues for the linear modes of the one dimensional slab}

Let us assume that the  perturbed quantities
depend on time through a factor  $\propto e^{-i\sigma t}$. In
this case the free oscillations of the gas column are described by
equation (\ref{e30}) for any value of $\sigma $ provided that we
substitute the factor $\sigma^{2}-\Omega^{2}$ for the factor $2\Omega
\omega $ in the second term on the left hand side.
 Clearly, the solutions to the problem can be found from
equations (\ref{e36}) and (\ref{e38}) with the parameter $\kappa $
expressed in terms of $\sigma $ as $\kappa =\sqrt {1+3{\sigma^{2}\over
\Omega^{2}}}$.

\subsubsection{The case of two free boundaries}
 If we assume that velocities are regular at both boundaries,  taken to be where
the density vanishes, this regularity condition determines a set of eigenvalues with
frequencies $\sigma_{n} \ge \Omega$. The set of $\sigma_{n}$
as well as the associated eigen functions  can be
easily determined from  (\ref{e38}).

It is important for our discussion that the mode corresponding to the
smallest eigenvalue, $\sigma=\Omega,$ has a  velocity
that is independent of $\zeta.$
\footnote{The regular modes with $\sigma_{n} > \Omega $ are just
 standing sound waves.}. Therefore
 this mode describes a uniform translation of the column as a
whole.

\subsubsection{The case of an upper rigid boundary}
When we adopt the boundary condition (\ref{e75})
together with the regular boundary condition at $\zeta \rightarrow
-\zeta_{0},$ the eigen frequency of the shift mode acquires a small
correction $\Omega \rightarrow \Omega +\omega $. The resulting mode
may be called as a 'modified shift' mode.
As we mentioned above one can consider the column to be rotating around some
centre with angular frequency $\Omega $. In the non-rotating frame,
 the shift mode of the free system describes a stationary
 displacement of the column as a whole, while for the  case with a rigid boundary,
 the modified shift mode
corresponds an oscillation of the column with characteristic time
scale $\sim 1/\omega $.

The change to the eigenfrequency, $\omega, $ can be readily be
determined by applying the boundary condition (\ref{e75})
to the solution given by equation (\ref{e38}) under 
the assumption that $|a|$ and $\lambda =\rho_{*}/\rho_{c}\ll 1$.

This is  found to be approximately given by
\be \tilde \omega =-{(i-{a\over 5\lambda })\over (1 +{1\over
25}({a\over \lambda })^{2})}{4a\over 15\pi}, \label{e76}\ee
where we recall that $\tilde \omega =\omega /\Omega $.
It follows from (\ref{e76}) that
when $a > 0$ the perturbations of the column decay with time.
The condition $a < 0$ leads to instability. Making the identification
$\tilde \omega_{r}\equiv -{3\over 10}a$ we  see that equation
(\ref{e76}) is precisely equivalent to equation (\ref{e58}) with
$\tilde \omega_{\nu}=0.$ Therefore, the linear
dynamics of the gas column with boundary condition (\ref{e75}) should
capture the form of the vertical motions relevant to the disc dynamics.

We see from equation (\ref{e76}) that the real and imaginary parts
of $\tilde \omega$, namely $\omega_{R}$ and $\omega_{I}$ are functions of only one
variable $y=a/\lambda.$  Thus we have
\be {\omega_{R}\over \lambda}= {4\over 75\pi}{y^{2}\over (1 +{1\over
25}y^{2})}, \quad
{\omega_{I}\over\lambda}= -{4\over 15\pi}{y\over (1 +{1\over
25}y^{2})}. \label{e77}\ee
Thus, all models with different $a$ and $\lambda $ but the same ratio
of these quantities have the same value of $\omega $  expressed
in units of $\lambda $ . As seen from
equation (\ref{e77}) the absolute value of $\omega_{I}$ has a
maximum when $y=5$ with $|\omega_{I}|={2|a|\over 15\pi}.$
This corresponds to the condition (\ref{e74}).

\subsection{Non linear numerical solutions for the vertically stratified slab}\label{Nonlin}
In our numerical work we use the simple Lagrangian  staggered
leap frog  scheme
described in Richtmyer $\&$ Morton 1967 for an ideal gas with
$\gamma =5/3.$ We use 400 grid points in the  vertical direction uniformly distributed with respect to
the vertical coordinate in the static initial state. 
Artificial viscosity is used in order to handle
possible shocks and it is checked that our results are essentially
independent on the value of the   artificial viscosity coefficient.
In order to check the scheme we compared the eigenvalues for the free
problem $\sigma_{n}$ and the Rankine-Hugoniot relations for shocks
with what is obtained in numerical computations. In both cases we obtained good
agreement between  analytical and numerical approaches.

In our computations we use the dimensionless time $\tau=\Omega t. $
The density $\rho $,
pressure $p$, velocity $v$ and position of a gas element are assumed
to be expressed in units of $\rho_{c}$, $P_{c}$, $\Omega \zeta_{0}$
and $\zeta_{0}$, respectively.
These normalisations are used to determine the normalisations of
all other quantities that make them dimensionless.
For example the thermal energy per unit
area of the unperturbed disc, $E_{th}={2\over \gamma
  -1}\int^{\zeta_{0}}_{0}d\zeta p$. Following our procedure, this
quantity is expressed in units  of $P_{c}\zeta_{0}.$
Thus we have $E_{th}={5\pi \over 48}\approx 0.33$ for $\gamma =5/3$ in
these units.

\subsubsection{Numerical check of the eigenvalues and
structure of the linear eigenmodes }

In order to check the validity of the linear theory we consider the
case $a > 0$ corresponding to decay of warp like perturbations. The
case $a < 0$ leads to a model with unphysical behaviour 
because it is such that, not only the modified shift mode, but
also the higher order acoustic modes are unstable. The growth rates of the latter
 are always larger than that of the modified shift modes leading them
 to dominate the evolution.

For $a > 0,$ we start our computations with initial
state having the density and pressure distributions corresponding to
hydrostatic  equilibrium and a very small uniform vertical
velocity $v = 10^{-4}$. We evolve the system for a very long time
$\tau \sim 10^{3}-10^{4}.$  Then we determine power spectra for
the state variables and use them to locate the peak corresponding to the
modified shift mode. It is confirmed that the associated spectral feature or line has a
Lorentz profile. The location  of the maximum determines the real part of the eigen frequency
$\tilde \omega_{R}$, and the width of the line determines $\tilde
\omega_{I}$.

We consider four different values of $\lambda= 0.032,0.056, 0.084,
0.11$ and seven different values of $a$ for a given value of $\lambda,$
such that $a=80\lambda/2^{k},$ where the integer $k = 1,2 ..7.$
When $a$ is smaller than $5\lambda $ - the value
corresponding to the maximal value of $\omega_{I}$ we expect the
column dynamics to be similar to the disc dynamics in the regime
described by Pringle 1996. In this case the values of velocity
and the Lagrangian displacement should not depend significantly on
$\zeta $ and the column response to the external pressure is similar
to that of a rigid body.  However, when  $a > 5\lambda ,$
we expect the behaviour to be similar to the disc dynamics in the regime
described in section \ref{smalllam}. In this regime
the upper disc surface behaves as if it is in contact with rigid wall
while the regions beneath  participate in stable
oscillations. In this case velocities and displacements
should have a strong dependence on $\zeta $ near the upper surface of
the column. In a case of a very large $a$ velocities and displacements
should be close to zero at $\zeta \approx \zeta_{0}$.

\begin{figure}
\begin{center}
\includegraphics[width=16cm,angle=0]{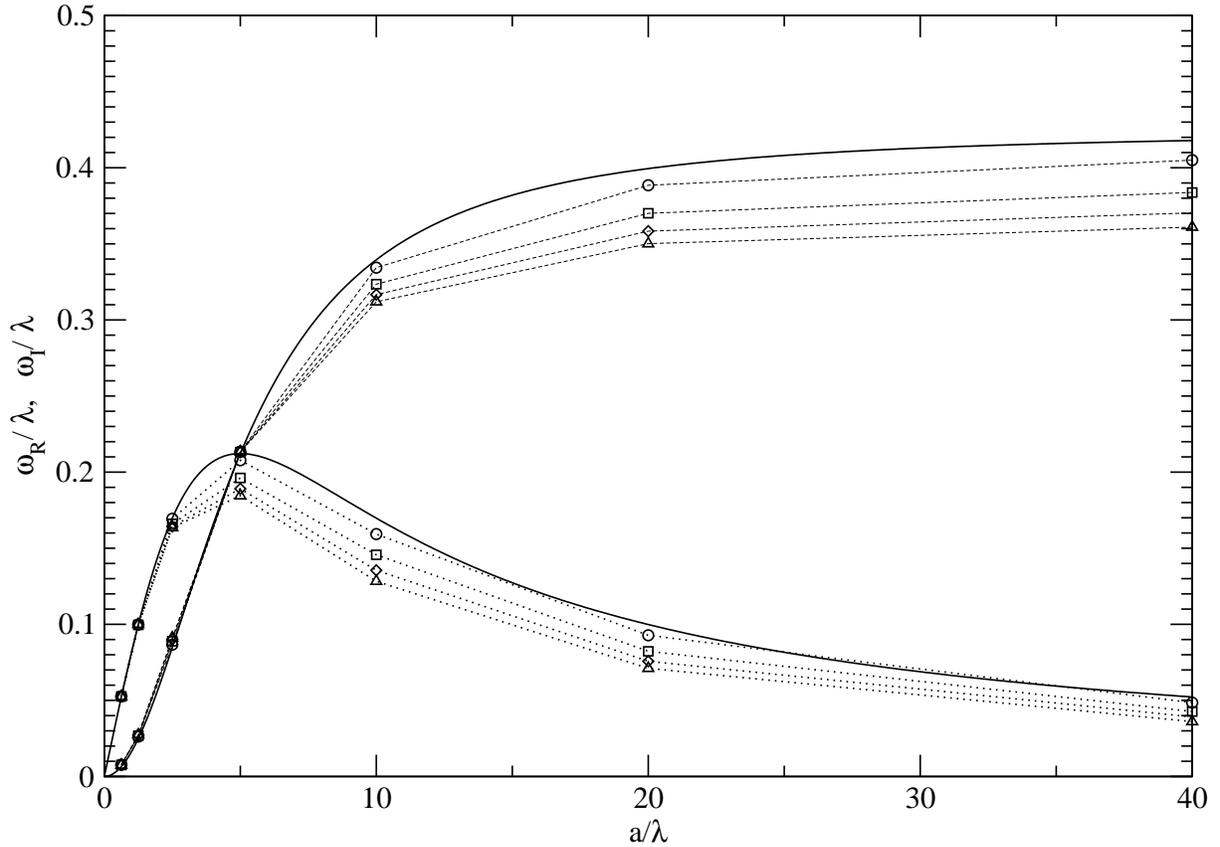}
\end{center}
\caption{Numerically obtained values of the real 
and imaginary parts of the eigen frequency, $\omega_{R}$ and $\omega_{I}$ divided by the
parameter $\lambda$ as functions of the ratio $a/\lambda $.  Solid
curves represent analytical results given by equation (\ref{e77}).
Dashed and dotted curves represent numerical values of $\omega_{R}$
and $\omega_{I}$, respectively.
The various symbol styles indicate different values of
$\lambda.$  These are such that smaller values of $\lambda $ lie
 closer to the analytically determined curves.}
\label{fig01}
\end{figure}

\begin{figure}
\begin{center}
\includegraphics[width=16cm,angle=0]{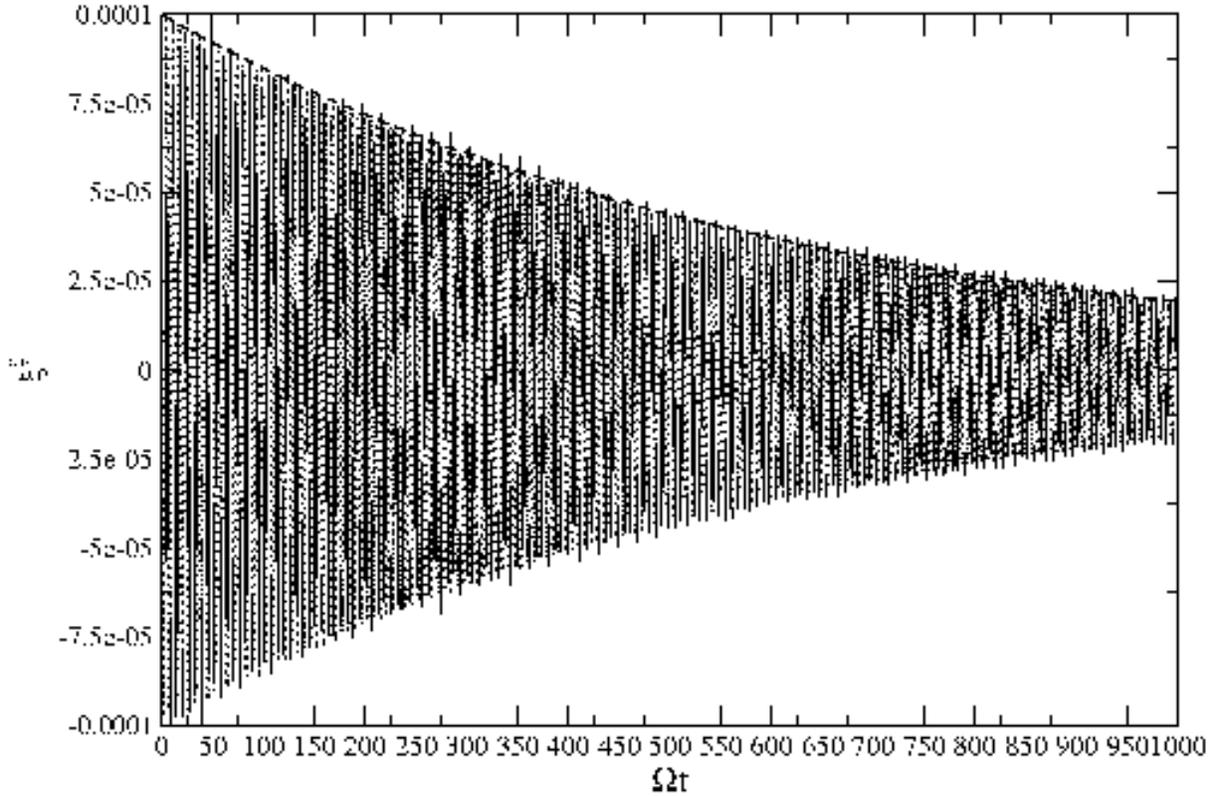}
\end{center}
\caption{The time dependence of the displacement vector evaluated near the
upper boundary of the gas column calculated for models with $\lambda
 =0.032$. The solid (dotted) curve corresponds to $a=1.26$
 ($a=0.02$). Note that both curves practically coincide.
  The dashed curve represents the exponential dependence of the
 amplitude of the oscillations on time, being  $\propto e^{-\omega_{I}t},$
 calculated with the analytically determined
 value of $\omega_{I}=1.65\cdot 10^{-3}$. }
\label{fig02}
\end{figure}

\begin{figure}
\begin{center}
\includegraphics[width=16cm,angle=0]{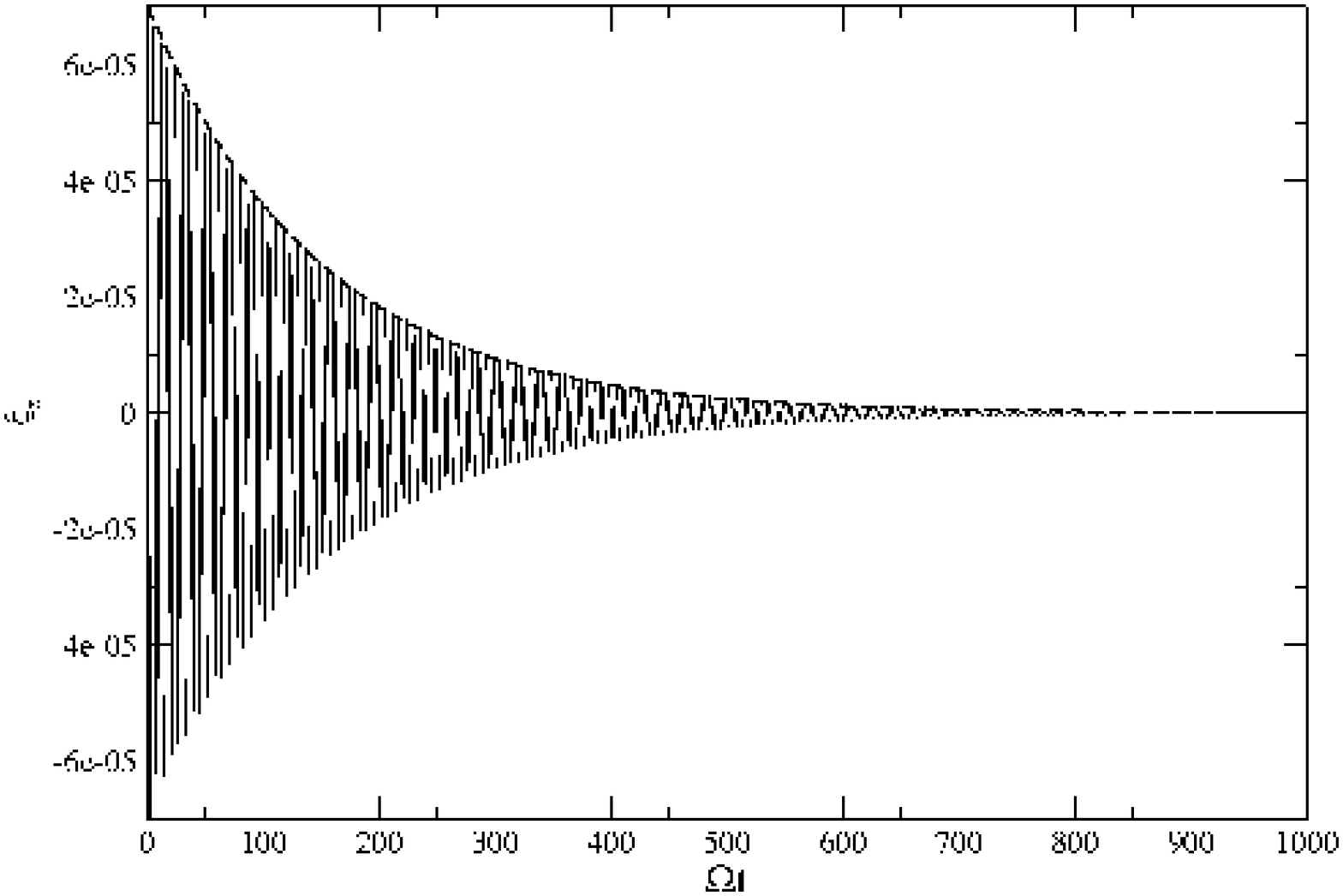}
\end{center}
\caption{ As Figure 2 but with $a=0.16$. This value of $a$
gives  the maximal decay rate when  $\lambda
 =0.032$.}
\label{fig03}
\end{figure}

The results of our computations are plotted in Figures 1-4. In Figure
1 we show the numerically obtained dependencies of $\omega_{R}/\lambda$ (the dashed
curves ) and   $\omega_{I}/\lambda$ (the dotted curves)
on $a/\lambda $ for the set of parameters described above.
Specific symbols are associated  with particular values of $\lambda. $
 Circles, squares,
diamonds and triangles represent the  specific values of $\lambda $ we used in
increasing order. Solid curves correspond to the analytically determined
dependencies given by equation (\ref{e77}). One can see that the
agreement between the analytic and numerical results is good
for the smallest value of $\lambda= 0.032.$ However, the curves
deviate more significantly as  $\lambda$ is increased.
This departure from a dependence
of $\omega/\lambda$  only on $a/\lambda$ may be explained as an effect due to the increasing importance of
of higher order terms in the
power series representation of $\omega/\lambda$  as a function of  $\lambda, $
for fixed $a/\lambda,$
that were not taken into account  in
equations (\ref{e43}) and (\ref{e43a}).

\begin{figure}
\begin{center}
\includegraphics[width=16cm,angle=0]{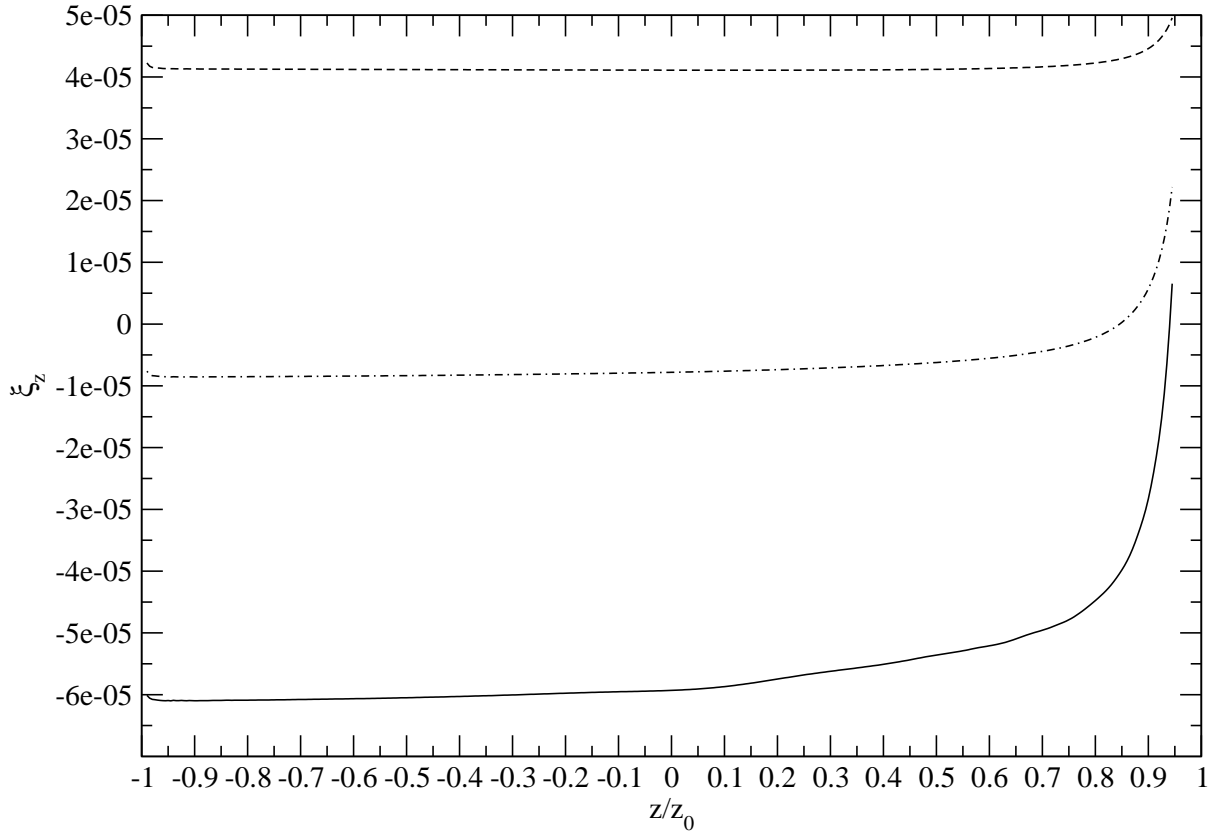}
\end{center}
\caption{The dependence of the displacement vector on the vertical
 coordinate $\zeta ,$  at a typical time, for $\lambda
 =0.032$ and different values of $a.$  The solid, dot-dashed and dashed
 curves correspond to $a=1.26, 0.16$ and $0.02$ respectively.}
\label{fig04}
\end{figure}

Figure 2 shows  the displacement vector
evaluated near the boundary $\zeta=\zeta_{0}$ as a function of time
for   $\lambda= 0.032$  with
$a=1.26$ (solid curve ) and $a=0.02$ ( dotted
curve). Regardless of the fact that these values of $a$ differ,
 their analytically calculated
decay rates are practically the same and such that
$\omega_{I}\approx 1.65\cdot 10^{-3}$. As expected
the temporal  behaviour of the displacement vector is almost
indistinguishable for these two cases. The dashed curve indicates the
exponential dependence on time, calculated in this case, using the decay rate
$\omega_{I}= 1.65\cdot 10^{-3}.$ It is evident that agreement
between the analytical and numerical results is  good.

Figure 3 is similar to Figure 2 but the dependence of the displacement
vector on time is shown for $a=0.16.$. Because $a/\lambda =5$
 this value of $a$ corresponds to the maximal decay rate
when $\lambda= 0.032.$  In this case it is given by $\omega_{I}\approx 6.7\cdot 10^{-3}.$
Note again the very good agreement between the
theoretical and numerical results.

In Figure 4 we show the dependence of the displacement vector on $\zeta $
calculated at the same  time
for  $a=1.26$ -  solid curve,
$a=0.16$ -  dot-dashed curve and $a=0.02$ -dashed curve. One
can see from this Figure that the displacement corresponding to the small
value of $a$ is practically independent of $\zeta.$ But
 for the large value of $a$ it has a prominent
increase toward the upper boundary of the column.
 Also, note that
in the latter case the value of displacement vector near the upper
boundary is close to zero. Thus, although the time dependence of
the displacement vectors corresponding to the cases of large and small
$a$ is practically the same, the spacial dependence of
these quantities is quite different.

\subsubsection{Non linear calculations}

\begin{table*}
 \centering
 \begin{minipage}{140mm}
  \caption{ \label{table1} The initial amplitude of
velocity profiles, $C$, the maximal value of the relative pressure difference
$ (P-P_0)/P_0 \equiv \Delta P/P_{0}$ calculated for the first pulsational period and
the difference between the final and initial values of the thermal
energy per unit area,  $\Delta E_{Th}$, for the different models used in the computations.
Note that quantities are expressed as multiples of the dimensional units
defined in the text.}
  \begin{tabular}{@{}llrrrrlrlr@{}}
  \hline
   $ $ & $1$ & $2$ & $3$ & $4$ & $5$ & $6$ & $7$ & $8$\\

 \hline
   $C$ & $10^{-3}$  &   $5\cdot 10^{-3}$  & $10^{-2}$ & $2\cdot 10^{-2}$
   & $4\cdot 10^{-2}$ & $8\cdot 10^{-2}$ & $1.6\cdot 10^{-1}$ &
   $3.2\cdot 10^{-1}$ \\
   $\Delta P/P_{0}$ & $4\cdot 10^{-2}$  & $0.2$  & $0.4$ & $1$ & $2.5
   $ & $7$ & $20 $ & $80 $ \\
   $\Delta E_{Th}$ &  $9\cdot 10^{-8}$ &  $1.4\cdot 10^{-6}$ &  $5.4\cdot
   10^{-6}$ & $2\cdot 10^{-5}$ &
 $8.6\cdot 10^{-5}$ & $4.4\cdot 10^{-4}$ &  $2\cdot 10^{-3}$ & $9\cdot 10^{-3}$  \\

\hline
\end{tabular}
\end{minipage}
\end{table*}

\begin{figure}
\begin{center}
\includegraphics[width=16cm,angle=0]{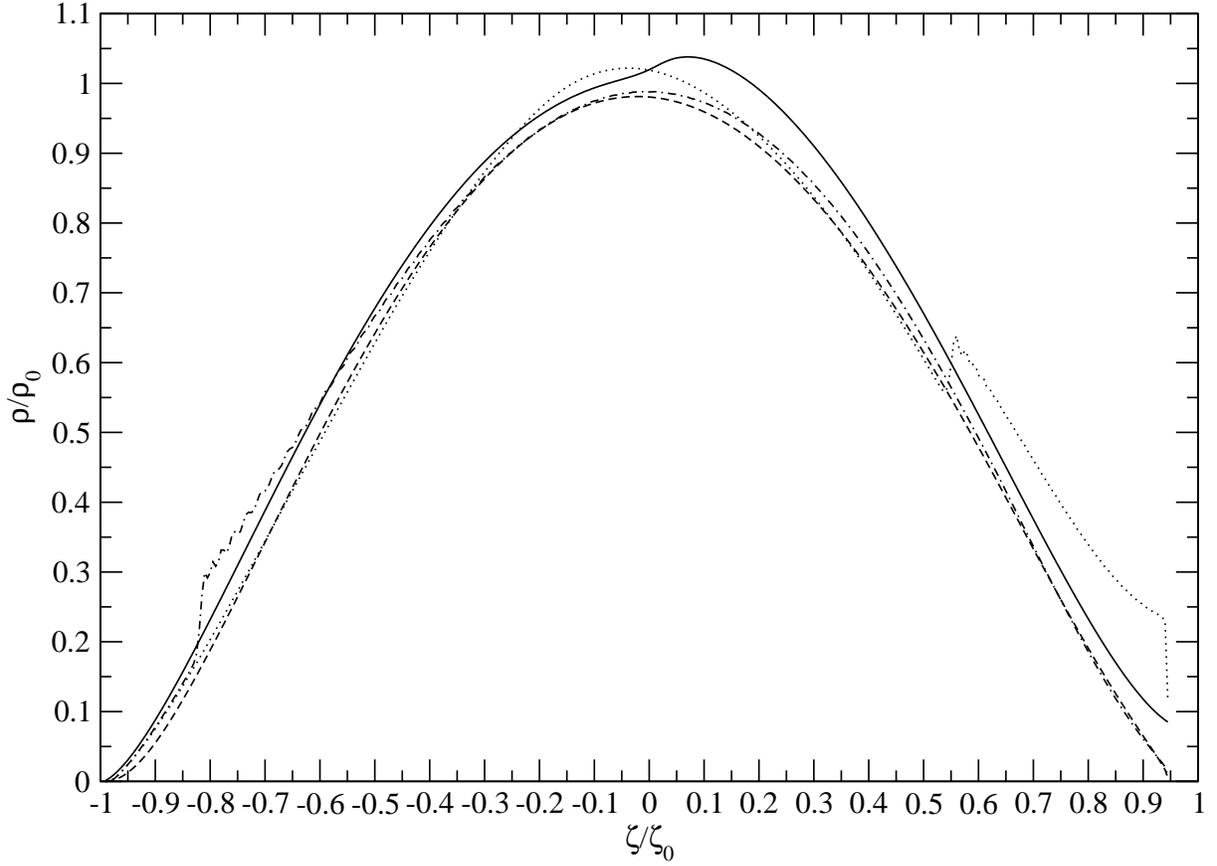}
\end{center}
\caption{The dependence of density $\rho $ on  $\zeta $ at
four different  times for model $7,$ which is characterised by the
relative pressure amplitude $C=0.16$. The solid, dashed, dotted and dash-dotted curves
correspond to the  times $\tau=2.5, 5, 7.5$ and $10$, respectively.}
\label{fig05}
\end{figure}

\begin{figure}
\begin{center}
\includegraphics[width=16cm,angle=0]{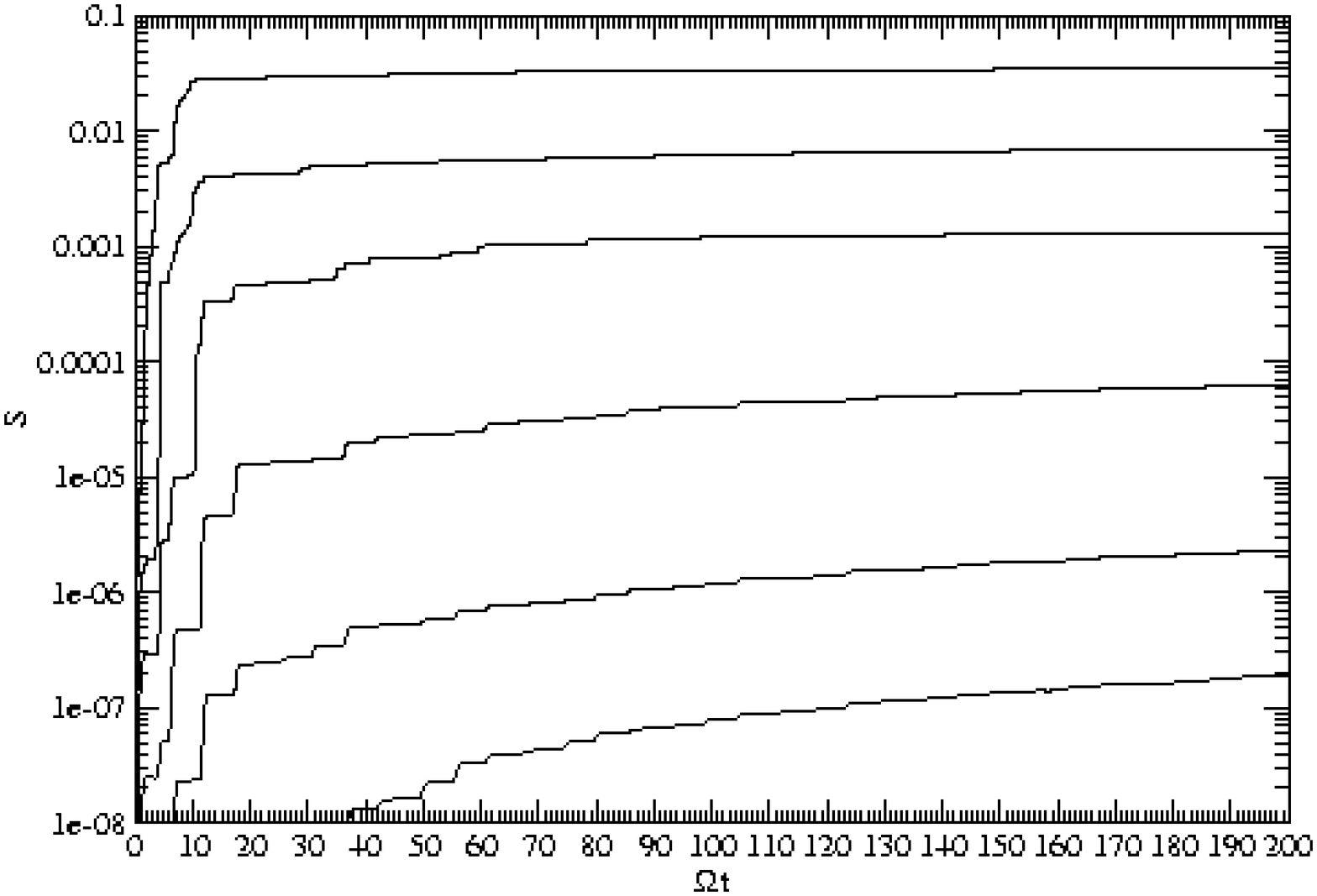}
\end{center}
\caption{The dependence of the entropy per unit area, $S,$ for
the six  models $(3-8).$  As the entropy increases while moving
from the lowermost to uppermost curve, the  models range from $3-8$
 with increasing values of the initial
amplitude $C$.}
\label{fig06}
\end{figure}

In order to determine how our results depend on the
initial velocity amplitude,  we consider the case
for which  the upper boundary of the column is a rigid wall, resulting
in the condition  that $v(\zeta
=\zeta_{*})=0$. From our previous discussion it follows that this
occurs when the parameter $a$ is sufficiently large. The uniform initial
velocity profile used in our previous numerical calculations is not
compatible with this boundary condition. Therefore, we use a
different  initial form for the velocity
\be v = C\cos ({\pi\over 2}{\zeta \over \zeta_{*}}). \label{e78}\ee
Since the non-linear effects are expected to operate on a short
dynamical time scale of order of a few periods $P=2\pi\Omega^{-1}$
we follow the evolution for relatively small times $\tau_{end}=200.$

When the rigid wall boundary condition is applied and the linear theory is
 valid, the column dynamics is determined
by the linear superposition of strictly  periodic oscillations.
Departures from this type of motion are
determined by non-linear effects. One of the most important for our purposes
 is the possibility of shock formation near the upper boundary.
When such shocks propagate through the column they irreversibly heat up the
gas thus causing dissipation of the motion induced in the column.
\footnote{Of course, shock formation in a realistic system
would be quite different from what we consider in the text. Since the
upper boundary of the disc is close to the surface of unit optical
depth, in a realistic situation, the shocks would be
radiative. However, it is plausible
that our study indicates the range of disc parameters for which
shocks may be expected.}

The strength of non-linear effects is obviously proportional to the
induced amplitude $C.$ In the first row of Table 1 we show the eight
values of $C$ used in our computations and the corresponding models, enumerated by numbers
from one to eight. In all cases we assume that the rigid boundary is
situated at the value of $\zeta_{*}$ corresponding to $\lambda = 0.032$.

A useful measure of non-linearity of the system is
the ratio of the  difference between the pressure
at the upper boundary and its equilibrium value $\Delta P=P-P_{0},$ to
$P_{0}$. For definiteness we estimate the maximal values of this
quantity during the first non-linear pulsation and display them in the second
row of Table 1. As seen from this Table
when $\Delta P/P_{0} < 1,$ this quantity has an approximate linear
dependence on $C,$ while for larger values this dependence is
somewhat steeper. The third row represents the difference between
the thermal energy (per unit area) at the end of the computations
 at $\tau_{end}=200$ and its initial value $E_{th}\approx 0.33$, in units
of $P_{c}\zeta_{0}$. As expected this quantity is approximately
proportional to the square of $C$.

\subsection{Shock development and the formation of a rarefied hot atmosphere}\label{8.3}
When $\Delta P/P_{0} < 1$ the dynamics of the model is well
described by the linear theory. In the opposite case,  non-linear
effects are substantial. The system exhibits a complicated pulsational
behaviour and strong shocks are observed. This regime is illustrated in
Figure 5 where the results of calculations for model $7,$ with
large initial amplitude $C=0.16,$ are presented.  We show the  density
profiles  at four different  times $\tau=2.5 $ ( solid
curve ), $\tau=5 $ (dashed curve ), $\tau=7.5 $ ( dotted  curve
) and $\tau=10 $ ( dot dashed curve ). The density distribution
at $\tau=2.5 $ is quite similar to the equilibrium density distribution.
When  $\tau=5 $ and $\tau=7.5 $ shocks are observed
propagating from the upper boundary in the direction of negative
$\zeta $. These shocks heat the gas up, thus dissipating  kinetic
energy. Also note the  density drop  at the upper boundary
when $\tau =5$ and $\tau =10$. We have checked that
this density drop has a persistent character and the density at the
boundary at $\tau=\tau_{end}$ is significantly smaller than the
initial density while the pressure does not exhibit
such a significant change.

Thus, the presence of shocks in the system
leads to formation of a hot rarefied region near the rigid boundary.
It is not clear however, how this effect would operate in
twisted discs. In the latter case the boundary is defined by the
condition that the optical thickness is one. The position of this
boundary may move when any density redistribution becomes significant.
In our case with fixed upper boundary location,
the presence of this hot low density region leads to the suppression of shock formation
after a time  corresponding to a few
orbital periods, an effect that may not occur if the upper boundary moved
in such a way as to maintain the relative pressure variations.

The effect in our case is illustrated in Figure 6 where we show
the dependence of entropy per unit area,
$S=\int d\zeta \rho \log (p/\rho^{\gamma})$ as a function of time
for the models with indices larger than $2$, in the natural units
$\rho_{c}\zeta_{0}.$  Curves attaining larger values
correspond to larger amplitudes. One can see from this Figure that
when the initial amplitude is sufficiently large, i.e.  $C \ge 0.04$ and
accordingly  $\Delta P/P_{0} \ge 2.5,$ half of the entropy change is
roughly determined for $\tau < 10-15.$
(i.e. approximately during the first two orbital periods). Then
shock production becomes less efficient due to the mechanism discussed
above.

\subsection{ Estimate of dissipation time scale}\label{8.4}
As mentioned above, it is not clear to what extent
the dynamics of the column in the strongly non linear regime is similar
to the non-linear dynamics of  twisted discs. One can assume, e.g.
that in  twisted discs, the boundary where  external radiation pressure applies
follows a certain density level. In this situation it seems reasonable
to assume that the shock formation might persist at a rate similar
to what is observed at initial times $\tau < 10$. In this
case the disc perturbations could be damped on a time scale $t_{nl}$
which can be roughly estimated from the results presented in Figure
1.

For model $7,$ assuming the thermal energy production is
of the order of $\Delta E_{th}=0.002,$  we can estimate the non linear
decay time as $t_{nl} \sim 10(E_{th}/\Delta E_{th})\Omega^{-1}\sim
1.5\cdot 10^{3}\Omega^{-1} $ where we recall that $E_{th}\approx 0.3$
in our units. Thus, under these assumptions the
perturbations may decay in approximately $200$ orbital periods.
This rate is characteristically the internal energy content within one
surface scale height per orbital period and becomes noticeable when the
relative pressure variation $\Delta P/P_0$  near the upper boundary
exceeds $\sim 2.5.$

Let us stress, however, that although indicative, because of an obvious sensitive
dependence on the density structure near the boundary,  such an estimate cannot be rigorously justified
within our model. Clearly  more comprehensive future investigations are needed.

\section{ Discussion and Conclusions}\label{Conc}

In this paper we have presented a self-consistent calculation
of the response of a twisted disc to the action of a radiation pressure force acting in
upper layers of the disc. This is assumed to be due to the interception  
and subsequent reradiation  of radiation from  a central source.
The  radiation pressure force is assumed to be  applied at a
single density level $\rho_{*}$ corresponding to optical thickness unity.
The degree of twist and warping is assumed to be small enough
that linear theory can be used to calculate the response.

The  analysis of  Pringle
1996  modelled the disc as a set of concentric
rigid rings in a state of Keplerian rotation. These were assumed
to communicate with each other through the 
 exchange angular momentum because of the action of viscous forces.
Up to now there has been no consideration of 
effects arising from the response of the disc vertical structure
to the externally applied pressure.

In this paper we have considered the warping and twisting of an accretion disc
taking account the response of the disc vertical structure to an external
 radiation field assuming that the
effect of self-shielding of radiation by the global twisted disc is not
significant. We developed a description of the evolution of the disc in terms
of a pair of equations 
 governing   the one dimensional 
evolution of the inclination as a function of  radius and time
(see section \ref{sec6} and equations (\ref{e26}), (\ref{e48}) and (\ref{e49})).
This description extends earlier ones, so that  the influence
of radiation pressure on  discs for which warps  occur in 
 the low  viscosity
bending wave regime, as well as the higher viscosity diffusive
regime,  may be considered.
 
We found several qualitatively different
dynamical regimes that may be realised in astrophysical discs.
These are related to whether the character of the response of the disc vertical
structure causes significant departures from what would be obtained for a rigid body. 

\subsection{ Conditions for  a quasi-rigid response}
 We found that to avoid such departures, the momentum flux  due to 
radiation from the central source $F_0,$  should be  smaller than a
critical value, $F_{*},$  given by 
$F_{*}=(\lambda /\delta )P_{c},$ where $\delta = \zeta_0/r $ is the disc 
 aspect ratio, $P_{c}$ is the
disc mid plane pressure and it has been assumed that 
 the the ratio $(\lambda /\delta )$ is  small (see sections \ref{Dyneq} and \ref{7.1.2}). 
Then, when $F_0 > F_{crit}\approx 0.1(\delta /\alpha )P_{c},$
warping instability of the disc becomes a possibility ( eg. Pringle, 1996). 

Thus for radiation driven warping
of a disc that behaves like an ensemble of rigid rings, 
 we find two requirements, namely  that  $F_{*} > F_{crit},$  together with 
$F_0 < F _{*}.$
These conditions together imply that the ration of the 
surface to mid plane density  
should be 
sufficiently large and such that 
 $\lambda > \lambda_{crit}\approx 0.1\delta^{2}/\alpha.$

\subsection{ Large external radiation momentum flux}
In the opposite limit of large  $F_0 > max(F_{*}, F_{crit}),$
consideration of the disc
structure in the vertical direction is essential as the 
vertical displacement ceases to be uniform.
The perturbed gas motion in the disc is
mainly determined by the vertical component of velocity and the
perturbed quantities tend to oscillate at a characteristic
frequency $\omega = (\rho_c \zeta_0 / 2\Sigma)\lambda \Omega $, with $\Omega$
 being the local Keplerian angular velocity.
In this situation, vertical motion tends to be suppressed by the
external pressure field on the irradiated side of the disc while  warping motions persist in 
the bulk of the disc and  
on the shielded side of the disc (see sections \ref{Dyneq} and \ref{smalllam}).

The disc surface intercepting
the radiation tends to align parallel to the radiation rays, thus
decreasing the amount of intercepted radiation, while the
inclination angles associated with the disc mid plane and the
opposite free boundary of the disc oscillate, being approximately
equal to each other. In this limit the upper
surface of optical thickness one plays the role of an impenetrable
wall. 
Thus the development of instability of the inclination
angle due to radiation pressure effects (e.g. Pringle
1996) would be inhibited in this limit.

In principle, the presence of warping motions in this regime would 
be associated with variability on  a large
characteristic time scale $T_{ch}\sim \lambda^{-1}T_{K}$, where
$T_{K}$ and $\lambda $ are some 'typical' values of the orbital
period and the density ratio.

\subsection{Limits on the WKB growth rate}
Following Pringle 1996 we have considered  
possible instabilities using a WKB approach. As this neglects
potentially important global effects and boundary conditions
results are not definitive, nonetheless
they should give a good indication of when instability 
could be possible.

We find that when $F_{0} < F_{*}$ the 
growth rate increases with  
 $F_{0}$ while in the formal limit
$F_{0} \rightarrow \infty$ it approaches zero.
 Thus  in the linear theory there
is an upper limit for the growth rate of the instability of the
inclination angle which is always $\le (\rho_c \zeta_0/ \Sigma)\lambda\Omega $ 
in the absence of viscosity (see section \ref{7.3}).
When present, viscosity acts to damp any instability at a rate $0.1\delta^2\Omega /\alpha,$
leading again to the condition
 $\lambda > \lambda_{crit}\approx 0.1\delta^{2}/\alpha$
 for radiation warping instability to be feasible. 

\subsection{Conditions for non-linearity}
 Our results described above 
rely upon the applicability of linear perturbation theory.
As discussed in section \ref{7.5},  
the breakdown of linear theory is expected when 
the  Lagrangian change of pressure induced in the surface layers 
becomes  of the  order of the unperturbed 
pressure even though the change of inclination angle  could be very small.
The corresponding constraints on the inclination
angle are especially strong for disc models with a significant
drop of temperature toward the boundaries of the disc.

\subsection{Non linear calculations for the one dimensional slab analogue}

A direct numerical approach to the issues discussed above
is not yet feasible due to three-dimensional nature of the
problem and  the many physical processes involved. However, when
vertical motions dominate, the problem becomes
analogous to the one dimensional problem of vertical motions induced  in a
stratified gas column orbiting around a central source with
Keplerian angular velocity.

As discussed in Section \ref{8} the one dimensional
slab gives the same fundamental 
period of oscillation as obtained
from  the full disc theory  
when an appropriate 
boundary condition  on the upper surface of 
the column is specified. 
The dependence of the
eigenfrequency on theoretical parameters as well as the 
effective   
presence, in the limit of small $\lambda,$
of an impenetrable wall at the upper surface 
of the column
have been checked numerically.  Where they can be compared, 
agreement between our analytical and numerical
results is very good.

We also used the one dimensional model in order to investigate
possible consequences of non-linear behaviour in the system.
To do this we studied the motion of the slab
ensuing from the imposition of a vertical velocity profile 
with varying amplitude.
We found that when
the ratio of the Lagrangian pressure perturbation to the local
value of the pressure at the upper boundary of the disc, $\Delta P/P_{*}$,  becomes
larger than $1-10,$ strong shocks propagating downward into the column
are observed (see section \ref{8.3}). In principle, these shocks may lead to non-linear
dissipation of energy stored in the vertical motion 
and in section \ref{8.4} we made a very crude estimate of a possible
time scale of $200$ orbits  for $\Delta P/P_{*}\sim 10$. However, this
result must be checked in framework of a more sophisticated
numerical approach which takes into account physical processes
neglected in this study, most notably the effects of radiation
transfer in the upper layers of the column.

\subsection{Issues for future consideration}

In a fully  self-consistent study the dynamical
 effects of radiation pressure must
be studied together with effects determined by the  radiation heating
of the disc atmosphere. This can be done in  the most convenient way
within the  framework of the one dimensional model discussed above.

In a realistic disc model, where effects due to the  flaring of the disc
lead to the interception  and scattering of radiation in the disc
photosphere,  when axisymmetric and unperturbed,
radiation heating can significantly influence on or even
determine the value of the density ratio $\lambda.$
 
This parameter, being the ratio of the density 
at the absorbing surface to the mid plane density
 defines the boundary between different dynamical
regimes for a twisted disc. In this connection it is interesting
to note that in certain vertical models of X-ray heated accretion
discs, the ratio $\lambda $ can be quite small, being of order of
$10^{-4}-10^{-5}$, e.g. Jimenez-Garate et al 2002. In such
models the new dynamical effects discussed in this Paper would
certainly play an important role.

\section*{Acknowledgements }
\noindent
We are grateful to A. M. Beloborodov and G. I. Ogilvie for discussions.
PBI has been supported in part by RFBR grant 07-02-00886.


\begin{appendix}

\section{Expression for radiation pressure acting on the disc surface}\label{A}

Expressions for the radiation pressure acting on a surface element
of a twisted disc (or, alternatively an expression for the
corresponding torque acting on a disc annulus of radius $r$) have
been derived by a number of authors, e.g. Petterson 1977b,
Pringle 1996, Ogilvie $\&$ Dubus 2001, under different
approximations. However, these authors assume that the disc
consists of rigid rings of different radii having differing
inclinations and orientations with respect to a fixed Cartesian
coordinate system, when deriving the corresponding expressions.
This assumption is essentially equivalent to the statement that
all vertical motions can be accounted for by a rigid tilt and it
precludes internal vertical motions that vary with the vertical
coordinate.

As discussed in the main text, this is not strictly valid, as
in general such motions occur. Accordingly,
the density surfaces  corresponding to an optical thicknesses of order
of unity where the radiation pressure force is applied are not everywhere parallel to the
surface locally corresponding to the disc mid-plane and the expression
for the radiation pressure acting on the disc must be,  accordingly  modified.

In general, an expression describing the surfaces of constant optical
thickness can be easily obtained from the transformation law
from the Cartesian coordinate system to the twisted coordinate
system written for a gas element situated near the surface of the disc
\be x=r\cos \phi \pm \beta \zeta_{0}\sin \gamma, \quad
y=r\sin \phi \mp \beta \zeta_{0}\cos \gamma ,  \label{ea1} \ee
and
\be z= \pm \zeta_{0} +\xi^{\zeta}(\zeta \rightarrow \zeta_{0})
+r\beta \sin(\phi -\gamma), \label{ea2}\ee
where the upper (lower) sign corresponds to the upper (lower)
disc surface. Let us recall that we assume that the inclination angle
$\beta $ is small and derive all equations in the linear approximation
for  the angle $\beta $ and the Lagrangian displacement vector $\bar \xi$.

The unit vector perpendicular to the disc surface is given by the standard
expression
\be \bar { n } = {\partial \bar {R }\over \partial r}\times
{\partial\bar  { R }
\over \partial \phi}{\rm {\huge {/}}} \left|{\partial \bar {R }\over \partial r}\times
{\partial\bar { R }
\over \partial \phi}\right|,  \label{ea3}\ee
where $\bar {R}$ is the radius vector with components $(x,y,z)$.

We make the standard assumption that the radiation propagates radially
before being intercepted by the disc and neglect the effect of self-shielding.
In this case an amount of radiation intercepted by the disc and subsequently
reradiated away is proportional to cosine of angle between the radius
vector and the normal to the disc, $\cos \Xi \equiv \bar { { n }}\cdot
\bar {{R } }/|\bar {{R} }|$. For the upper branch of the surface
a simple calculation gives
\be \cos \Xi =-r{\partial \over \partial r} {(\zeta_{0} +
\xi^{\zeta})\over r}-r{\cal } {\cal W}, \label{ea4}\ee
where we recall that ${\cal W}=\beta^{'}\sin \psi -\beta
\gamma^{'}\cos \psi = \Psi_{1}\sin \phi -\Psi_{2}\cos \phi$. As we
discuss in the main text, in this paper, for simplicity, we neglect
the effect of disc flaring and accordingly set ${\partial \over
\partial r} ({\zeta_{0}\over r})=0$ in equation (\ref{ea4}).

The upper (lower ) branch of the disc surface is able to intercept the
radiation of a central source only when $\cos \Xi  < 0$
( $\cos \Xi  >  0$).  Therefore the radiation pressure, $F$, applied to
the disc surface has the form
\be F_{\pm}=\mp \Theta (\mp \cos(\Xi )) {2\over 3} F_{0} \cos
(\Xi), \label{ea5}\ee
where $\Theta (x)$ is the step function,
the signs $(+)$ and $(-)$ stand for the upper and lower branches
of the disc surface, respectively,
\be F_{0}={L\over 4\pi c r^{2}}  \label{ea5a}\ee
is the momentum flux per
unit of surface carried by the radiation emitted by the central
source, and $L $ is the luminosity of the source.
The factor $(2/3)$ accounts for (assumed ) isotropic character of the
disc reradiation.

In order to have  the same symmetries as the surface pressure term
(\ref{ea5}) the $\zeta $-component of the displacement vector,
$\xi^{\zeta}$, and, accordingly, the $\zeta$-component of velocity
perturbation, $v^{\zeta}$, must satisfy
\be \xi^{\zeta}(\zeta, \phi)= -\xi^{\zeta}(-\zeta, \phi +\pi), \quad
v^{\zeta}(\zeta, \phi)= -v^{\zeta}(-\zeta, \phi +\pi). \label{ea6}\ee
On the other hand, for a given value of $\phi$, $v^{\zeta}$ may be
separated on even and odd components:  $v^{\zeta}=v_{e}^{\zeta}+v_{o}^{\zeta}$
and both components should be taken into
account in order to satisfy the appropriate boundary conditions
at $\zeta \rightarrow \pm \zeta_{0}$. However, the latter
requirement poses a problem with the symmetry  law (\ref{ea6}).
Indeed, while the even part of $v^{\zeta} $, $v_{e}^{\zeta}$,
agrees with (\ref{ea6}) the odd one is not
and the standard separation of the velocity
perturbations into even and odd parts
although being respected by the perturbed equations
of motion, is not, strictly speaking, compatible with
the symmetries provided by the boundary conditions. This problem
can be circumvented with the help of the assumption that the
normal decomposition into even and odd parts in $\zeta$ of the perturbed
vertical velocity is replaced by
\be v^{\zeta} =v^{\zeta}_{e} -
sign (\sin(\phi-\phi_1)) v^{\zeta}_{o}
, \label{ea7}\ee
where $v^{\zeta}_{e}$ and  $v^{\zeta}_{o}$
 have even (odd ) properties with respect to
the reflection $\zeta \rightarrow -\zeta $. Since the surface radiation pressure
vanishes when $\cos\Xi =0$ or equivalently
$\sin(\phi -\phi_1) =0$ and, accordingly, we have $v^{\zeta}_o=0$ at the
corresponding values of $\phi =\phi_{1}$ and $\phi=\phi_{1}+\pi$
the velocity anzatz (\ref{ea7}) is self-consistent.
Also, it agrees with the symmetries determined by
equation ({\ref{ea5}}) and it has local simmetries respected by
the equations of motion.

Above and in the main text we always consider the angle
$\phi $ to be in a segment $[\phi_{1}..\phi_{1}+\pi]$
defined by the condition $\sin (\phi- \phi_1 ) > 0. $
With help of this
convention we can treat our dynamical variables as having the
standard even and odd symmetries with respect to reflection
$\zeta \rightarrow -\zeta.$
Continuation of these variables into
the segment defined by condition $\sin (\phi-\phi_1 )< 0 $ can be made
with the help of the expression (\ref{ea7}). In the case when
$\sin (\phi- \phi_1 ) > 0 $ it follows from equations (\ref{ea4}) and
(\ref{ea5}) that we have
\be F(\rho=\rho_{*}, \zeta \rightarrow \zeta_{0})\equiv F_{+}={2\over 3}F_{0}r{d
\over d r}\left({\xi^{\zeta}\left(\rho_{*}\right)\over r}+{\cal W}\right), \label{ea8}\ee
and
\be F(\rho=\rho_*,\zeta \rightarrow -\zeta_{0})\equiv F_{-}= 0, \label{ea9}\ee
where we recall that the density $\rho_{*}$ corresponds to the surface
of an unperturbed disc with optical thickness $\tau \approx 1.$
Taking into account that we can write $F_{+}=F_{+}^{1}\cos \phi +
F_{+}^{2}\sin \phi $ and $\xi^{\zeta}=\xi_{1}^{\zeta}\cos \phi +
\xi_{2}^{\zeta}\sin \phi $, we can represent equation (\ref{ea8}) in a
convenient complex form
\be {\bf F}_{+}={2\over 3}F_{0}r{d
\over d r}\left({\bf \xi}^{\zeta}/ r+i{\bf W}\right ),
\label{ea10}\ee
where ${\bf F}_{+}=F_{+}^{1}+iF_{+}^{2}$,
${{\xi}}_{\zeta}=\xi_{1}^{\zeta}+i\xi_{2}^{\zeta}$
and ${\bf W}= \beta e^{i\gamma}=\Psi_{1} +i \Psi_{2}$. Obviously, we
also have ${\bf F}_{-}=0$.

\section {Solution of equation (45) for the case of arbitrary
$\gamma $}\label{B}

When the parameter $\gamma >1$ is not specified, the homogeneous equation (\ref{e30})
can be reduced to a simpler form with help of equations (\ref{e3}) and
(\ref{e4}). We get
\be (1-x^{2}){d^{2}\over d x^{2}}{\bf v}-{2\gamma x\over \gamma -1}
{d\over d x}{\bf v}+{4\over \gamma -1}\tilde \omega {\bf v} =0, \label{eb1}\ee
where $x=\xi/\xi_{0}$. A general solution of (\ref{eb1}) has the form
\be {\bf v}=(1-x^{2})^{-{1\over \gamma -1}}\left(C_{1}F({1\over 2}-\alpha,
{1\over 2}-\beta, {1\over 2}, x^{2})+C_{2}xF(1-\alpha,
1-\beta, {3\over 2}, x^{2})\right), \label{eb2}\ee
where $F(\alpha, \beta, \delta, x)$ is the hyper-geometric function and
\be \alpha, \beta ={\gamma +1\over 4(\gamma -1)}(1\pm
\sqrt{1+{64(\gamma -1)\over (\gamma +1)^{2}}\tilde \omega
}). \label{eb3}\ee.
When $\tilde \omega \ll 1$ we have
\be \alpha \approx {\gamma +1\over 2(\gamma -1)}+{8\over \gamma
  +1}\tilde \omega, \quad \beta \approx -{8\over \gamma
  +1}\tilde \omega. \label{eb4}\ee

\section{The gauge transformations}\label{C}

As we pointed out in the main text different twisted coordinate
systems can describe the same physical situation provided that they
are connected by a certain family of transformations (gauge
transformations) which maintain fixed distributions of physical
quantities characterising the system (density and
velocity fields in the case of a baratropic gas)
in a non rotating (say, cylindrical) coordinate system
with origin located at the central source. These  transformation laws can
be found by considering transformations from the cylindrical to the
twisted coordinates. We derive them here assuming, for simplicity, that
the equation of state is baratropic and the gas is inviscid. The
latter assumption allows us to neglect the presence of the radial
component of velocity in an unperturbed background state.

We begin by considering the transformation of density.
Let the density $\rho $ be represented in the cylindrical and twisted
coordinate systems as
\be \rho=\rho_{0}(r_{c},z_{c})+\rho_{1}^{c}(r_{c},z_{c})=\rho_{0}(r,\zeta)+\rho_{1}(r,\zeta),
\label{ec1}\ee
where the background part $\rho_{0}$ has the same functional
dependence on the cylindrical and twisted coordinates.  The
cylindrical and twisted coordinates are related to each other by
transformations (\ref{e8a})-(\ref{e8c}). Substituting these
transformations in equation (\ref{ec1}) we obtain
\be\rho_{1}^{c}(r, \zeta ) = \rho_{1}(r, \zeta)+D\beta \sin (\phi
-\gamma ), \label{ec2}                   \ee
where
\be D=\left(\zeta {\partial \over \partial r}-r{\partial \over \partial
  \zeta}\right)\rho_{0}. \label{ec3}\ee
By construction the quantity $\rho_{1}^{c}$ is not changed after a
gauge transformation. Therefore, it may be classified as a 'gauge
independent' quantity. The result (\ref{ec2}) can be represented in a
compact form with help of the complex notation introduced in equation (\ref{e9})
\be {\bf \rho}_{1}^{c}={\bf \rho}_{1}+iD{\bf W}. \label{ec3a}\ee

In an absolutely analogous way we can find the law of transformation of velocity perturbations.
We have
\be {\bf v}^{\phi_{c}}={\bf v}^{\phi}+\zeta (i(r\Omega)^{'}{\bf W}-\dot
{\bf W}), \label{ec4}\ee
\be {\bf v}^{z_{c}}={\bf v}^{\zeta}+ir\dot {\bf W}+r\Omega
{\bf W}, \label{ec5}\ee
and
\be {\bf v}^{r_{c}}={\bf v}^{r}-i\zeta \dot {\bf W}-\zeta \Omega
{\bf W}. \label{ec6}\ee
Here the quantities ${\bf v}^{\phi_{c}}$, ${\bf v}^{z_{c}}$
and ${\bf v}^{r_{c}}$ represent the velocity perturbations in the
cylindrical coordinate system. They are gauge independent by
construction. The quantities ${\bf v}^{\phi}$, ${\bf v}^{\zeta}$
and ${\bf v}^{r}$ represent the time derivative of the
corresponding coordinates of a comoving  gas element. As we discussed above
these quantities do not coincide with the projections of
velocity onto the orthonormal basis used in the formalism exploiting
the twisted coordinate system.
The relations between these projections and the time derivatives of
the coordinates can be easily found from the results provided
in the Appendix of Petterson 1978, his
equations (A3). They can be written in the form
\be \hat {\bf v}^{\phi}={\bf v}^{\phi}-\zeta \dot {\bf W}, \quad
\hat {\bf v}^{\zeta}={\bf v}^{\zeta}+ir \dot {\bf W}, \quad
\hat {\bf v}^{r}={\bf v}^{r}-i\zeta \dot {\bf W},
\label{ec7}\ee
where it is implied that the background parts of
the $(\zeta)$ and $(r)$-components of velocity can be neglected.
Substituting equations (\ref{ec7}) in equations (\ref{ec4})-(\ref{ec6})
we find
\be {\bf v}^{\phi_{c}}=\hat {\bf v}^{\phi}+i\zeta(r\Omega)^{'}{\bf W}, \label{ec8}\ee
\be {\bf v}^{z_{c}}=\hat {\bf v}^{\zeta}+r\Omega
{\bf W}, \label{ec9}\ee
and
\be {\bf v}^{r_{c}}=\hat {\bf v}^{r}-\zeta \Omega
{\bf W}. \label{ec10}\ee

Now let us suppose that there are two twisted coordinate systems
with two different ${\bf W}_{1}$ and ${\bf W}_{2}$
describing the same physical system. That means that the
quantities characterising the density perturbation and perturbations
of the velocity field in the fixed cylindrical system must be the same
when expressed in terms of Euler angles and corresponding perturbations
belonging to the different twisted coordinate systems. From this
requirement and equations
(\ref{ec3a}) and (\ref{ec8})-(\ref{ec10}) it follows that the
differences in the corresponding to these systems perturbations of
density and velocity components must be proportional to the difference
between  ${\bf W}_{1}$ and ${\bf W}_{2}$, $\delta {\bf W}={\bf W}_{1}-{\bf W}_{2}$,
\be \delta {\bf \rho}_{1}=-iD\delta {\bf W}, \label{ec11}\ee
\be \delta \hat {\bf v}^{\phi}=-i\zeta(r\Omega)^{'}\delta {\bf W}, \label{ec12}\ee
\be \delta \hat {\bf v}^{\zeta}=-r\Omega \delta {\bf W}, \label{ec13}\ee
and
\be \delta \hat {\bf v}^{r}=\zeta \Omega
\delta {\bf W}. \label{ec14}\ee
Equations (\ref{ec11})-(\ref{ec14}) define the gauge transformations
between two twisted coordinate system for an inviscid disc.
Viewed in these terms, the
generalisation to the case of a viscid disc is straightforward
provided that the background part of the $(r)$-component of velocity is
taken into account.

One can also  check that equations (\ref{ec11})-(\ref{ec14}) provide an
exact solution of the equations of motion and the continuity equation
for an inviscid disc being written in a twisted coordinate system,
for an arbitrary dependence of ${\bf W}$ on $t$ and $r$. This solution
describes the so-called trivial modes corresponding to an
unperturbed disc. In this case the density and velocity fields in the
cylindrical coordinate system remain equal to their unperturbed values
while the dynamics in the twisted coordinates is such that perturbations of
density and velocity defined with respect to the twisted coordinates
are exactly compensated by those associated with time dependent Euler angles.

It is important to stress that our dynamical equations (\ref{e11a}) and
(\ref{e12a}) as well as the continuity equation (\ref{e15}) are not
invariant under the transformations generated by (\ref{ec11})-(\ref{ec14}) even
when the term determined by viscosity are discarded. This is due to
the fact that certain terms are neglected in these equations. These
terms provide only next order corrections in our expansion series in
small parameters   when the gauge (\ref{e40}) is fixed. On the other
hand  the combination ${\bf \xi}^{\zeta}/ r+i{\bf W}$ entering in
the expression (\ref{ea10}) determine an invariant quantity - the
angle between the normal to the disc surface and the radial direction.
This combination must be the same in all twisted coordinate systems
connected by transformations (\ref{ec11})-(\ref{ec14}). Using
equations (\ref{e21a}) and (\ref{ec7}) it is easy to see that this
combination is proportional to
\be \hat {\bf v}^{\zeta}+r\Omega {\bf W}. \label{ec15}\ee
From equation (\ref{ec13}) it follows that the expression (\ref{ec15})
is, indeed, a gauge invariant quantity.

\end{appendix}

{}

\clearpage

\centerline {\bf        }
\vspace{1 cm}

\clearpage

\end{document}